\documentclass[usenatbib]{mn2e}

\IfFileExists{srcltx.sty}{\usepackage[active]{srcltx}}

\usepackage[normalem]{ulem}

\usepackage{amsmath,amssymb,xspace} %
\usepackage{graphicx,color,paralist,times} %

\newcommand{\keV}{\ensuremath{\:\mathrm{keV}}} %
\newcommand{\kev}{\ensuremath{\:\mathrm{keV}}} %
\newcommand{\cm}{\:\mathrm{cm}} 
\newcommand{\s}{\:\mathrm{s}} 
\newcommand{\pc}{\:\mathrm{pc}} 
\newcommand{\kpc}{\:\mathrm{kpc}} 
\newcommand{\dm}{{\textsc{dm}}} 
\newcommand{\xmm}{\textsl{XMM-Newton}\xspace} 
\newcommand{\fov}{\mathrm{fov}} 
\newcommand{\for}{\mathrm{Forn}}
\newcommand{\scl}{\mathrm{Scul}}
\renewcommand{\S}{\mathcal{S}} 

 \newcommand{\ks}{\mathrm{ks}} %
\newcommand\Chandra{%
  {\it Chandra}}

\newcommand\tabref[1]{%
Tab.~\ref{tab:#1}}

\newcommand\figref[1]{%
  Fig.~\ref{fig:#1}}

\usepackage{url}

\begin{document}

\title[Searching for dark matter in X-rays]{Searching for dark matter in
  X-rays: how to check the dark matter origin of a spectral feature} %

\author[Boyarsky et al.]{%
  Alexey~Boyarsky$^{1,2}$, Oleg~Ruchayskiy$^1$, Dmytro~Iakubovskyi$^2$,
  \newauthor{Matthew G.~Walker$^3$, Signe~Riemer--S{\o}rensen$^4$, Steen~H.~Hansen$^4$}\\
  $^1$Ecole Polytechnique F\'ed\'erale de Lausanne, FSB/ITP/LPPC, BSP
  CH-1015, Lausanne, Switzerland\\
  $^2$Bogolyubov Institute for Theoretical Physics, Metrologichna str.,
  14-b, Kiev 03680, Ukraine\\
  $^3$Institute of Astronomy, University of Cambridge, UK\\
  $^4$Dark Cosmology Centre, Niels Bohr Institute, University of Copenhagen,
  Juliane Maries Vej 30, DK-2100 Copenhagen, Denmark.\thanks{The Dark
    Cosmology Centre is funded by the Danish National Research Foundation.}}

\date{}

\maketitle

\begin{abstract}
  A signal from decaying dark matter (DM) can be unambiguously distinguished
  from spectral features of astrophysical or instrumental origin by studying
  its spatial distribution. We demonstrate this approach by examining the
  recent claim of \protect\citet{Loewenstein:09} regarding the possible DM
  origin of the 2.5 keV line in Chandra observations of the Milky Way
  satellite known as Willman~1.  Our conservative strategy is to adopt, among
  reasonable mass estimates derived here and in the literature, a relatively
  large dark mass for Willman~1 and relatively small dark masses for the
  comparison objects.  In light of the large uncertainty in the actual dark
  matter content of Willman~1, this strategy provides minimum exclusion limits
  on the DM origin of the reported signal.  We analyze archival observations
  by {\it XMM-Newton} of M31 and Fornax dwarf spheroidal galaxy (dSph) and
  {\it Chandra} observations of Sculptor dSph. By performing a conservative
  analysis of X-ray spectra, we show the absence of a DM decay line with
  parameters consistent with those of \protect\citet{Loewenstein:09}. For M31,
  the observations of the regions between 10 and 20 kpc from the center, where
  the uncertainties in the DM distribution are minimal, make a strong
  exclusion at the level above $10\sigma$. The Fornax dSph provides a $\sim
  3.3\sigma$ exclusion instead of a predicted $4\sigma$ detection, and the
  Sculptor dSph provides a $3\sigma$ exclusion instead of a predicted
  $2.5\sigma$ detection.  The observations of the central region of M31 (1-3
  kpc off-center) are inconsistent with having a DM decay line at more than
  $20\sigma$ if one takes the most conservative among the best physically
  motivated models. The minimal estimate for the amount of DM in the central
  40 kpc of M31 is provided by the model of \citet{Corbelli:09}, assuming the
  stellar disk's mass to light ratio $\sim 8$ and almost constant DM density
  within a core of 28 kpc.  Even in this case one gets an exclusion at
  $5.7\sigma$ from central region of M31, whereas modeling \emph{all}
  processed data from M31 and Fornax produces more than $14\sigma$ exclusion.
  Therefore, despite possible systematic uncertainties, we exclude the
  possibility that the spectral feature at $\sim 2.5$~keV found in
  \protect\citet{Loewenstein:09} is a DM decay line. We conclude, however,
  that the search for DM decay line, although demanding prolonged (up to
  1~Msec) observations of well-studied dSphs, M31 outskirts and other similar
  objects, is rather promising, as the nature of a possible signal can be
  checked. An (expected) non-observation of a DM decay signal in the planned
  observations of Willman~1 should not discourage further dedicated
  observations.
\end{abstract}

\section{Introduction}
\label{sec:introduction}

Modern astrophysical and cosmological data strongly indicate that a
significant amount of matter in the Universe exists in the form of \emph{dark
  matter}.  The nature of dark matter (DM) remains completely unknown and its
existence presents one of the major challenges to modern physics. It is
commonly believed that the dark matter is made of particles.\footnote{Although
  more exotic possibilities (such as, for example, primordial black
  holes,~\cite{Carr:05}) exist.}  However, the Standard Model of elementary
particles fails to provide a DM candidate. Therefore the hypothesis of a
particle nature of DM implies extension of the Standard Model.

\emph{Weakly interacting massive particles} (WIMPs) are probably the most
popular class of DM candidates. They appear in many extensions of the Standard
Model \citep[see e.g.][]{Bertone:04,Hooper:09a}.  These stable particles
interact with the SM sector with roughly electroweak strength~\citep{Lee:77}.
To provide a correct dark matter abundance, WIMPs should have mass in the GeV
range. A significant experimental effort is devoted to the detection of the
interaction of WIMPs in the Galaxy's DM halo with laboratory nucleons
(\emph{direct detection experiments}, see e.g.~\cite{Baudis:07}) or to finding
their annihilation signal in space (\emph{indirect detection experiments}, see
e.g.~\cite{Carr:06,Bergstrom:07a,Hooper:08}).

Another large class of DM candidates are \emph{superweakly} interacting
particles (so called \emph{super-WIMPs}), i.e.  particles, whose interaction
strength with the SM particles is much more feeble than the weak one.
Super-WIMPs appear in many extensions of the Standard Model: extensions of the
SM by right-handed neutrinos \citep{Dodelson:93,Asaka:05a,Lattanzi:07},
supersymmetric
theories~\citep{Takayama:00,Buchmuller:07,Feng:2004zu,Feng:2003xh}, models
with extra dimensions~\citep{Feng:2003xh} and string-motivated
models~\citep{Conlon:07}.  The feeble interaction strength makes the
laboratory detection of super-WIMPs challenging (see e.g.~\cite{Bezrukov:06}).
On the other hand, many super-WIMP particles possess a \emph{2-body radiative
  decay} channel: ${DM} \to \gamma + \nu,\gamma+\gamma$, producing a photon
with energy $E_\gamma = M_{DM}/2$. One can therefore search for the presence of
such a monochromatic line with energy in the spectra of DM-containing
objects~\citep{Abazajian:01b,Dolgov:00}.  Although the lifetime of any
realistic decaying DM is much longer than the lifetime of the Universe (see
e.g.~\cite{Boyarsky:08b}), the huge amount of potentially decaying DM particles in a typical halo means a potentially detectable decay signal in the spectra of
DM-dominated objects.  Searching for such a line provides a major way of
\emph{detection of super-WIMP DM particles}.

The strategy of the search for decaying dark matter signal drastically differs
from its annihilation counterpart. Indeed, the \emph{decay} signal is
proportional to the \emph{DM column density} -- integral of a DM distribution
$\rho_\dm$ along the line of sight $\S=\int_{l.o.s.}\rho_{DM}(r)dr$ as opposed
to $\int_{l.o.s.}\rho^2_{DM}(r)dr$ in the case for annihilating DM. The DM
column density varies slowly with the mass of DM
objects~\citep{Boyarsky:06c,Boyarsky:09b} and as a result a vast variety of
astrophysical objects of different nature would produce a comparable decay
signal~\citep{Boyarsky:06c,Bertone:07,Boyarsky:09b}.  Therefore
\begin{inparaenum}[\em (a)]\item without sacrificing the expected signal one has a freedom of choosing 
  observational targets, avoiding complicated astrophysical backgrounds;
\item if a candidate line is found, its surface brightness profile may be
  measured (as it does not decay quickly away from the centers of the objects)
  and distinguished from astrophysical lines that usually decay in outskirts
  of galaxies and clusters. Moreover, any tentative detection in one object
  would imply a signal of certain (comparable) signal-to-noise ratio from a
  number of other objects. This can be checked and the signal can either be
  unambiguously confirmed to be that of DM decay origin or ruled out.
\end{inparaenum}
This allows to distinguish the decaying DM line from any possible
astrophysical background and therefore makes astrophysical search for the
decaying DM \emph{another type of a direct detection experiment}.

In this paper we demonstrate the power of this approach. We check the recent
conjecture by~\citeauthor{Loewenstein:09} \citep[\textbf{LK09} in what
follows]{Loewenstein:09}, who reported that a spectral feature at $2.51\pm
0.07$~keV with the flux $(3.53 \pm 1.95)\times
10^{-6}\;\mathrm{photons\, cm^{-2} s^{-1}}$ (all errors at $68\%$ confidence level) in
the spectrum of the Milky Way satellite known as Willman~1~\citep{Willman:05}
may be interpreted as a DM decay line.  According to LK09, the line with such
parameters is marginally consistent the restrictions on decaying dark matter
from some objects~(see e.g.~\citet{Loewenstein:08,Riemer:09a} or the
restrictions from the $7$~ks observation of Ursa Minor dwarf spheroidal
in~\citet{Boyarsky:06d}).  The results of other
works~\citep[e.g.][]{Watson:06,Abazajian:06b,Boyarsky:07a} are inconsistent at
$3\sigma$ level with having $100\%$ of decaying DM with the best-fit parameters of
LK09.  However, it is hard to exclude completely new unknown systematic
uncertainties in the DM mass estimates and therefore in expected signal, from
any given object. To this end, in this work we explicitly check for a presence
of a line with parameters, specified above, by using archival observations by
\xmm and \textit{Chandra} of several objects, where comparable or stronger
signal was expected. We compare the signals from several objects and show that
the spatial (angular) behavior of the spectral feature is inconsistent with
its DM origin so strongly, that systematic uncertainties cannot affect this
conclusion. We also discuss a possible origin of the spectral feature of LK09.

\section{Choice of observational targets}
\label{sec:choice-observ-targ}

The particle flux of the decaying dark matter (in photons\, cm$^{-2}\times{}$s))
is given by
\begin{equation}
  F_{DM}=\frac{\Gamma}{4\pi
    m_\dm}\frac{M_{\dm}^{\fov}}{D_{L}^{2}} = \frac{\Gamma}{4\pi
    m_\dm}\Omega_\fov \S,\label{eq:flux_distant}
\end{equation}
where $m_\dm$ is the mass of the dark matter particle, $\Gamma$ is the decay
rate to photons (depending on the particular model of decaying dark matter),
$M_\dm^{\fov}$ is the dark matter mass within the instrument's field-of-view
(FoV) $\Omega_\fov$ and $D_L$ is the luminous distance towards the
object.\footnote{Eq.~(\ref{eq:flux_distant}) is valid for all objects whose
  size is much smaller than $D_L$.}

\textbf{Willman~1 signal.}  In LK09 the signal was extracted from the central
$5'$ of the Willman~1 observation. Assuming the DM origin of this signal, the
observed flux $(3.53 \pm 1.95)\times 10^{-6}\;\mathrm{photons\, cm^{-2} s^{-1}}$
corresponds to $F_{\rm W1} = (4.50 \pm 2.5)\times
10^{-8}\;\mathrm{photons\, cm^{-2} s^{-1} arcmin^{-2}}$ flux per unit solid angle.  The
mass within the FoV of the observation of Willman~1 was estimated in LK09 to
be $M_{\rm W1}^{\fov} = 2\times 10^6 M_\odot$, using the best-fit values
from~\cite{Strigari:07a}. Taking the luminous distance to the object to be $D_L
= 38\kpc$, the average DM column density of Willman~1 is $\S_{\rm W1} \simeq
208.5\: M_\odot\pc^{-2}$ (see however discussion in the
Section~\ref{sec:dwarf-spher-galax} below). To check the presence of a DM
decay line, we should look for targets with comparable signal.

The signal-to-noise ratio for a weak line observed against a featureless
continuum is given by (see e.g. discussion in~\citet{Boyarsky:06f}):
\begin{equation}
  \label{eq:3}
  (S/N) \propto {\S}{\sqrt{t_\text{exp} A_\text{eff}
      \Omega_\fov \Delta E}}
\end{equation}
where $\S$ is the average DM column density within an instrument's
field-of-view $\Omega_\fov$, $t_\text{exp}$ is the exposure time,
$A_\text{eff}$ is an effective area of the detector and $\Delta E$ is the
spectral resolution (notice that the line in question has a much smaller
intrinsic width than the spectral resolution of \textsl{Chandra} and \xmm).
For the purpose of our estimate we consider the effective areas of \xmm MOS
cameras and \Chandra{} ACIS-I (as well as their spectral resolution) to be
approximately equal, and effective area of PN camera is 2 times
bigger.\footnote{See e.g.
  \url{http://xmm.esa.int/external/xmm_user_support/documentation/uhb/node32.html}
  and
  \url{http://cxc.harvard.edu/ccr/proceedings/07_proc/presentations/drake/}.}

\begin{figure}
  \centering
  \includegraphics[width=.7\linewidth,angle=-90]{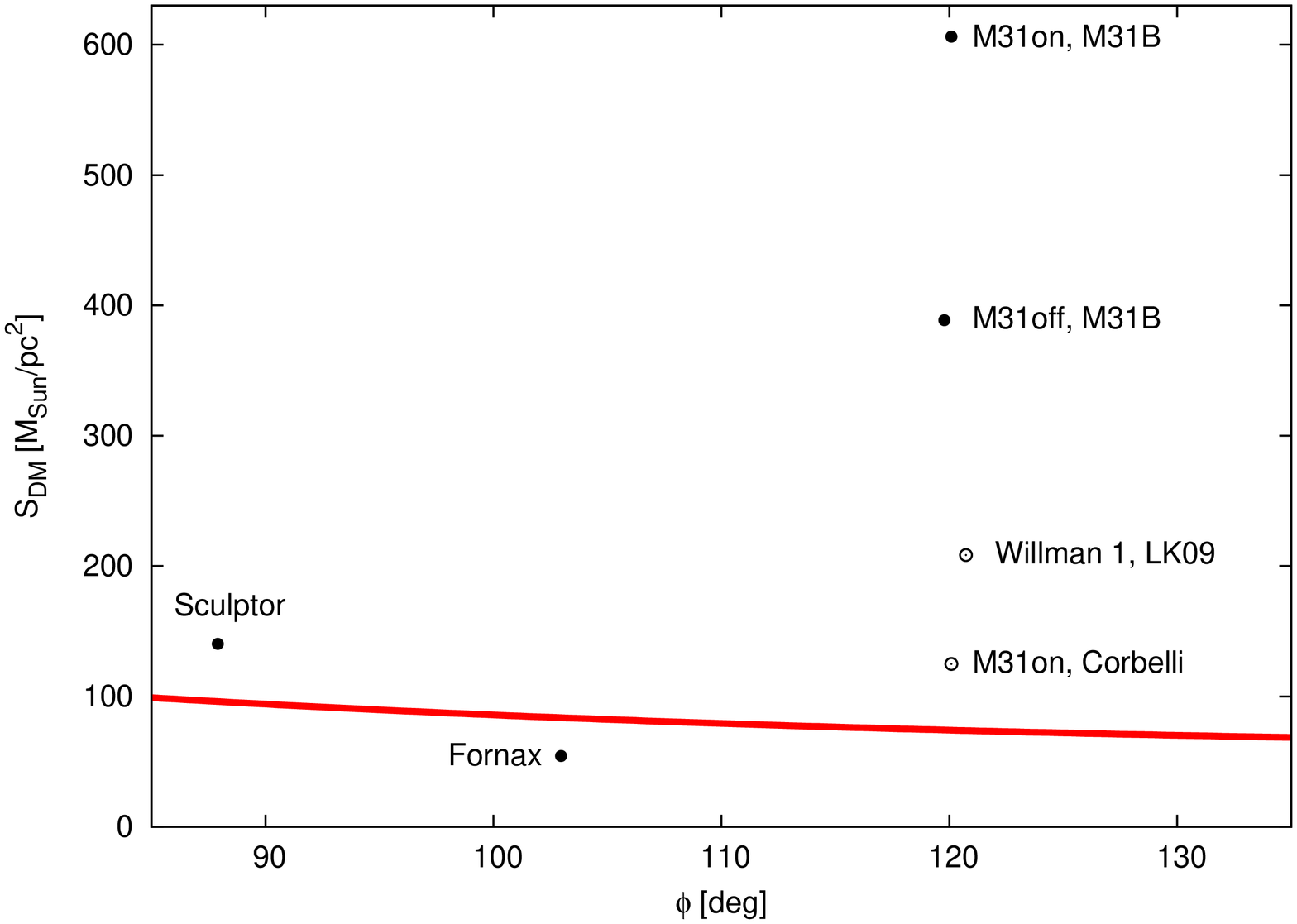}
  \caption{Comparison of DM column densities of the Milky Way (red solid line)
    with the density profiles of M31, Fornax and Sculptor dSphs, discussed in
    the text.  The DM column density estimate for Willman~1 is based on the
    ``optimistic'' DM density profile from \protect\citet{Strigari:07a}, used
    in LK09 (see Section~\ref{subsubsec:willman1} for discussion). The angle
    $\phi$ marks the direction off Galactic center (for an object with
    galactic coordinates $(l,b)$ $\cos\phi = \cos l \cos b$). }
  \label{fig:mw}
\end{figure}

\textbf{Galactic contribution.} The contribution of the Milky Way's DM halo
along the line of sight should also be taken into account, when estimating
the DM decay signal.\footnote{Notice, that both in LK09 and in this work
only instrumental (particle) background has been subtracted from the
observed diffuse spectra.} Using a pseudo-isothermal profile
\begin{equation}
  \rho_{\rm iso}(r) = \frac{\rho_c}{1+r^2/r_c^2},\label{eq:rho_iso}
\end{equation}
with parameters, adopted in~\cite{Boyarsky:06c,Boyarsky:06d} ($r_c = 4\kpc$,
$\rho_c = 33.5 \times 10^6 M_\odot/\kpc^3$, $r_\odot = 8 \kpc$), the
corresponding Milky Way column density in the direction of Willman~1
($120.7^\circ$ off Galactic center) is $73.9 M_\odot\pc^{-2}$. As demonstrated
in~\cite{Boyarsky:07b}, the value of DM column density computed with this DM
density profile coincides with the best-fit
Navarro-Frenk-White~\cite[\textbf{NFW}]{Navarro:96} profiles
of~\cite{Klypin:02} and \cite{Battaglia:05} within few $\%$ for the off-center angle
$\phi \gtrsim 90^\circ$.  Other dark matter distributions produce
systematically larger values of dark matter column density. For example, the
best-fit pseudo-isothermal profile of~\cite{Kerins:00} would corresponds to an
additional Milky Way contribution $\sim 119 M_\odot\pc^{-2}$ in the direction
of Willman~1.
The comparison between the DM column
densities of several objects with that of Milky Way in their directions is
shown in Fig.~\ref{fig:mw}.

\section{XMM-Newton data analysis}  %
\label{sec:data-analysis}

For each \xmm observation observation, we extracted both EPIC MOS and PN
spectra.\footnote{We use \texttt{SAS} version \texttt{9.0.0} and \texttt{XSPEC}
  \texttt{12.6.0}.}  We select events with \texttt{PATTERN <= 12} for MOS
and \texttt{PATTERN == 0} for PN camera and use \texttt{FLAG == 0}.  We used
\texttt{SAS} task \texttt{edetect\_chain} to exclude point sources, detected with the
likelihood values above 10 (about 4$\sigma$). For EPIC PN camera, bright
strips with out-of-time (OOT) events were also filtered out, following the
standard procedure.\footnote{As described for example at
  \url{http://xmm.esac.esa.int/sas/current/documentation/threads/EPIC_OoT.shtml}.}
We imposed stringent flare screening criteria.  To identify ``good time
intervals'' we performed a light-curve cleaning, by analyzing temporal
variability in the hard ($E>10\kev$) X-ray band (as described e.g.
in~\citealt{Read:03,Nevalainen:05,Carter:07}). The resulting light curves were
inspected ``by eye'' and additional screening of soft proton flares was
performed by employing a method similar to that described in
e.g.~\citet{Leccardi:08,Kuntz:08} (i.e. by rejecting periods when the count
rate deviated from a constant by more than $2\sigma$). The extracted spectra
were checked for the presence of residual soft proton contaminants by
comparing count rates in and out of
FoV~\citep{DeLuca:03,Leccardi:08}.\footnote{We use the script
  \url{http://xmm2.esac.esa.int/external/xmm_sw_cal/background/Fin_over_Fout}~\cite{DeLuca:03}
  provided by the \xmm\ EPIC Background working group.} In all spectra that we
extracted the ratio of count rates in and out of FoV did not exceed 1.13 which
indicates thorough soft proton flare cleaning \citep[c.f.][]{Leccardi:08}.

Following the procedure, described in~\cite{Carter:07} we extracted the
filter-wheel closed background\footnote{Available at
  \url{http://xmm2.esac.esa.int/external/xmm_sw_cal/background/filter_closed/index.shtml}}
from the same region as signal (in detector coordinates). The background was
further renormalized based on the $E>10$~keV band count rates of the extracted
spectra. The resulting source and background spectra are grouped by at least
50 counts per bin using \texttt{FTOOL}~\citep{ftools} command \texttt{grppha}.

\begin{table*}
\centering 
\begin{tabular}[c]{|c|c|c|c|}
  \hline ObsID & PL index & PL norm, $10^{-4}\mathrm{ph.\,
    cm^{-2}~s^{-1}~ \keV^{-1}})$ & $\chi^2$/dof\\
  &  &  at 1~keV (MOS1 / MOS2 / PN)\\
  \hline
  \multicolumn{4}{c}{\texttt{M31on} observations:}\\
  \hline
  \texttt{0109270101} & 1.33 & 5.71 / 5.80 / 7.88 & 1328/1363\\
  \hline
  \texttt{0112570101} & \ldots  & 5.37 / 6.12 / 6.86 & \ldots \\
  \hline
  \texttt{0112570401} & \ldots & 6.13 / 6.66 / 6.37 & \ldots\\
  \hline 
  \multicolumn{4}{c}{\texttt{M31off} observation:}\\
  \hline
  \texttt{0402560301} & 1.46 & 6.28 / 6.12 / 9.97 & 895/997\\
  \hline 
  \multicolumn{4}{c}{\texttt{M31out} observations:}\\
  \hline
  \texttt{0511380101} & 1.53 & 2.75 / 2.88 / 3.89 & 875/857\\
  \hline 
  \texttt{0109270401} & 1.47 & 5.02 / 4.41 / 5.22 & 606/605\\
  \hline 
  \texttt{0505760401} & 1.27 & 1.91 / 2.04 / 2.47 & 548/528\\
  \hline 
  \texttt{0505760501} & 1.28 & 2.25 / 1.85 / 3.03 & 533/506\\
  \hline 
  \texttt{0402561301} & 1.57 & 2.94 / 3.10 / 3.95 & 489/515\\
  \hline 
  \texttt{0402561401} & 1.47 & 4.33 / 3.58 / 4.47 & 751/730\\
  \hline 
  \texttt{0402560801} & 1.55 & 5.22 / 4.49 / 5.83 & 902/868\\
  \hline 
  \texttt{0402561501} & 1.44 & 2.70 / 4.29 / 6.67 & 814/839\\
  \hline 
  \texttt{0109270301} & 1.39 & 4.29 / 4.58 / 3.75 & 413/436\\
  \hline 
  \multicolumn{4}{c}{Fornax observation:}\\
  \hline
  \texttt{0302500101} & 1.52 & 4.46 / 4.02 / 3.78 & 938/981\\
  \hline
\end{tabular} 
\caption{Best-fit values of \texttt{powerlaw} index and normalization
  for each \xmm\ observation analyzed in this paper.  The systematic
  uncertainty is included (see text). In \texttt{M31on} region, \texttt{powerlaw} index for different observations is chosen to be the same, because they point to the same spatial region.}
\label{tab:best_fit_values} 
\end{table*} 

The extracted spectra with subtracted instrumental background were fitted by
the \texttt{powerlaw} model of \texttt{XSPEC} in the energy range 2.1--8.0~keV
data for MOS and 2.1--7.2~keV for the PN camera.  This choice of the
\emph{baseline model} is justified by the fact that for all considered objects
we expect no intrinsic X-ray emission above 2~keV and therefore the observed
signal is dominated by the extragalactic diffuse X-ray background (XRB) with
the powerlaw slope $\Gamma = 1.41\pm 0.06$ and the normalization $(9.8\pm1.0)
\times 10^{-7}~ \mathrm{photons\, cm^{-2} s^{-1} keV^{-1} arcmin^{-2}}$ at 1~keV (90\% CL
errors) (see
e.g.~\citealt{Lumb:02,DeLuca:03,Nevalainen:05,Hickox:06,Carter:07,Moretti:08}).
Galactic contribution (both in absorption and emission) is negligible at these
energies and we do not take it into account in what follows.  For all
considered observations \texttt{powerlaw} model provided a very good fit,
fully consistent with the above measurements of XRB.  The \texttt{powerlaw}
index was fixed to be the same for all cameras (MOS1, MOS2, PN), observing the
same spatial region, while the normalization was allowed to vary independently
to account for the possible off-axis calibration uncertainties between the
different cameras, see e.g.~\citealt{Mateos:09,Carter:09}.  The best-fit values
of the \texttt{powerlaw} index and normalizations are presented in
Table~\ref{tab:best_fit_values}.

To check for the presence of the LK09 line in the spectra, we added a narrow
($\Delta E/E \sim 10^{-3}$) Gaussian line (\texttt{XSPEC} model
\texttt{gaussian}) to the baseline \texttt{powerlaw} model. Assuming the DM
origin of this line, we fix its normalization at the level $F_{\rm W1}$,
derived in LK09, multiplied by the ratio of total (including Milky Way
contribution) DM column densities and instrument's field-of-view. We vary the
position of the line in the interval
corresponding to the $3\sigma$ interval of LK09: 2.30 -- 2.72 keV.
Eq.~(\ref{eq:3}) allows to estimate the expected significance of the line.

When searching for weak lines, one should also take into account uncertainties
arising from the inaccuracies in the calibration of the detector response and
gain. This can lead to systematic residuals caused by calibration
inaccuracies, at the level $\sim 5\%$ of the model flux (some of them having
edge-like or even line-like shapes) \citep[see
e.g.][]{CAL-TN-0018-2-0,Kirsch:04} as well as discussion
in~\cite{Boyarsky:06e} for similar uncertainties in \emph{Chandra}.  To
account for these uncertainties we perform the above procedure with the $5\%$
of the model flux added as a systematic error (using \texttt{XSPEC} command
\texttt{systematic}).

\section{Andromeda galaxy} %
\label{sec:andromeda-galaxy} 

\subsection{Dark matter content of M31}
\label{sec:dm-m31}

The Andromeda galaxy is the closest spiral galaxy to the Milky Way.  Its dark
matter content has been extensively studied over the years~\citep[see
e.g.][and references
therein]{Kerins:00,Klypin:02,Widrow:05,Geehan:06,Tempel:07,Chemin:09,Corbelli:09}
for an incomplete list of recent works.  The total dynamical mass (out to
$\sim 40$~kpc) can be determined from the rotation curve measured from {\sc
  Hi} kinematics.  The major uncertainty in determination of dark matter
content is then related to a separation of contributions of baryonic (stellar
bulge and especially, extended stellar disk) and dark components to the total
mass.\footnote{Some works also take into account a presence of supermassive
  black hole in the center of a galaxy~\citep[see e.g][]{Widrow:05} and
  additional contribution of a gaseous
  disk~\citep[c.f.][]{Chemin:09,Corbelli:09}. Their relative contributions to
  the mass at distances of interest turn out to be negligible and we do not
  discuss them in what follows.} The baryonic mass is often obtained from the
(deprojected) surface brightness profile (optical or infrared), assuming
certain mass-to-light ratio for the luminous matter in the bulge and the disk
of a galaxy.

\begin{figure}
  \includegraphics[width=\linewidth]{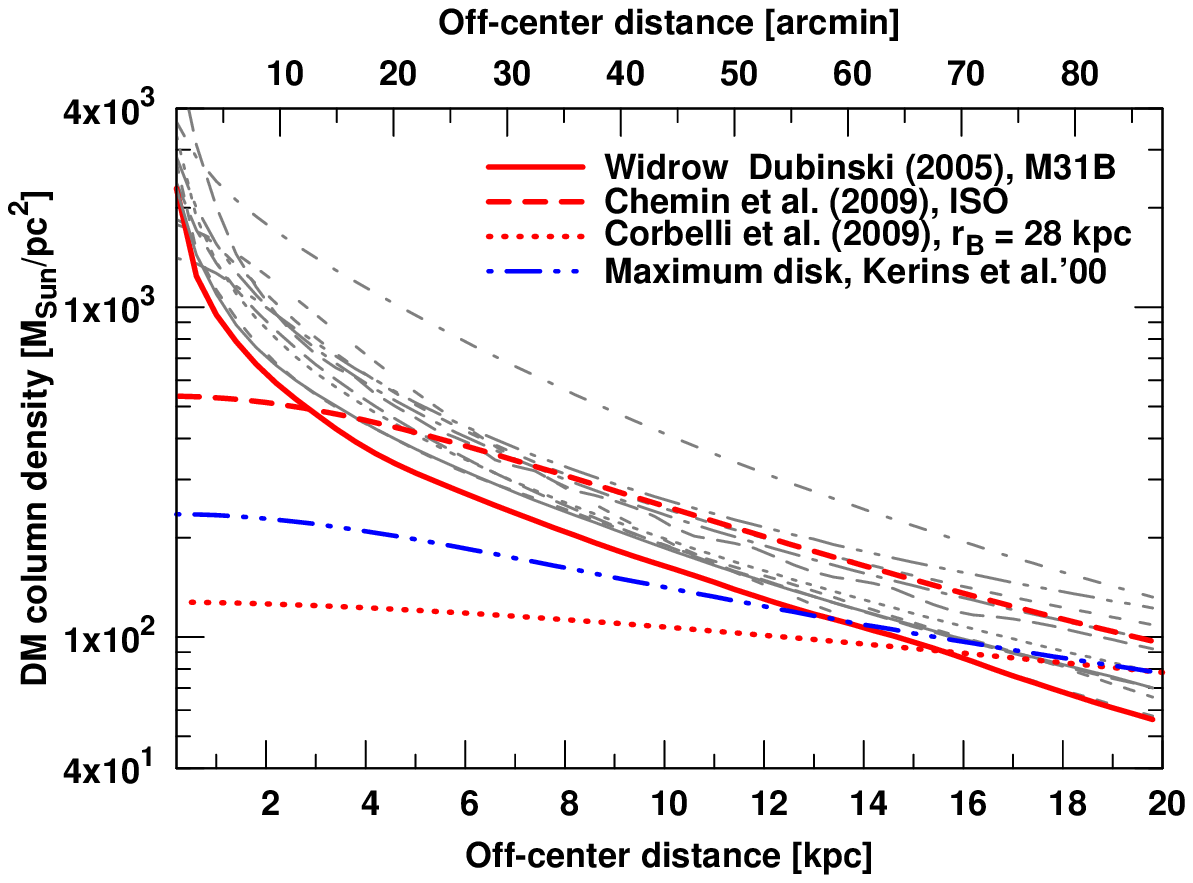}
  \caption{Dark matter column density of M31 as a function of off-center
    distance. The gray lines represent models from
    \protect\citep{Kerins:00,Klypin:02,Kerins:03,Widrow:05,Geehan:06,Tempel:07}
    analyzed in B08. The red solid line represents the most conservative model
    M31B of \protect\cite{Widrow:05}. The model of \protect\cite{Chemin:09}
    (red dashed line), the maximum disk model of \protect\cite{Kerins:03}
    (blue dashe double-dotted line) and the ``minimal'' model of
    \protect\cite{Corbelli:09} (red dotted line) (see text) are shown for
    comparison.}
\label{fig:DM_distrib-m31-2010}
\end{figure} 

\citet{Boyarsky:07a} (\textbf{B08} in what follows) analyzed dark matter
distributions of M31, existing in the literature at that time in order to
provide the most conservative estimate of the expected DM decay signal. The DM
column density $\S$ in more than 10 models from the
works~\citep{Kerins:00,Klypin:02,Widrow:05,Geehan:06,Tempel:07} turns out to
be consistent within a factor of $\sim 2$ for off-center distances greater
than $\sim 1$~kpc (see gray lines in Fig.\ref{fig:DM_distrib-m31-2010}). The
most conservative estimate of dark matter column density was provided by one
of the models of \citet{Widrow:05} (called in that work M31B), marked as red
solid line on Fig.\ref{fig:DM_distrib-m31-2010}.

Recently two new \textsc{Hi} surveys of the disk of M31 were
performed~\citep{Chemin:09,Corbelli:09} and new data on mass distribution of
M31 became available.\footnote{We thank A.~Kusenko and M.~Loewenstein in
  drawing our attention to these works~\citep{Kusenko:10a}.} In
\citet{Chemin:09} modeling of \textsc{Hi} rotation curve between $\sim
0.3$~kpc and 38~kpc was performed. \citet{Chemin:09} used \emph{R-band}
photometric information~\citep{Walterbos:87,Walterbos:88} to determine the
relative contribution of the stellar disk and the bulge.  Based on this
information and taking into account foreground and internal
extinction/reddening effects \citep[see section~8.2.2 in][for
details]{Chemin:09} they determined mass-to-light ratio, using stellar
population synthesis models~\citep{Bell:03} $\Upsilon_\text{disk} \simeq 1.7
\Upsilon_\odot$ (in agreement with those of~\cite{Widrow:05,Geehan:06}).
Aiming to reproduce the data in the inner several kpc, they explored different
disk-bulge decompositions with two values of the bulge's mass-to-light ratio
$\Upsilon_\text{bulge} \simeq 0.8 \Upsilon_\odot$ and $\Upsilon_\text{bulge}
\simeq 2.2 \Upsilon_\odot$ (all mass-to-light ratios are in solar units).
\citet{Chemin:09} analyzed several dark matter density profiles (NFW, Einasto,
cored profile). According to their 6 best-fit models (3 DM density profiles
times two choices of mass-to-light ratio) the DM column density in the
Andromeda galaxy is higher than the one, adapted in B08 (model M31B)
everywhere but inside the inner 1~kpc (see red dashed line on
Fig.\ref{fig:DM_distrib-m31-2010}).

\citet{Corbelli:09} used circular velocity data, inferred from the \textsc{Hi}
rotation curves, also extending out to $\sim 37\kpc$. The authors exclude from
the analysis the inner 8~kpc, noticing the presence of structures in the inner
region (such as a bar), associated with non-circular motions.
\citet{Corbelli:09} also use optical data of~\cite{Walterbos:88} and determine
an upper and lower bounds on the disk mass-to-light ratio using the same
models~\citep{Bell:00,Bell:03} as \citet{Chemin:09}. They obtain possible
range of values of the \emph{B-band} mass-to-light ratio $2.5\Upsilon_\odot
\le \Upsilon_\text{disk} \le 8\Upsilon_\odot$. These values do not take into
account corrections for the internal extinction.\footnote{The correction for
  foreground extinction was taken into account in~\citet{Corbelli:09}, using
  the results of~\citet{Seigar:06}. The uniform disk extinction was not
  applied (which explains high values of $\Upsilon_\text{disk}$). The authors
  of \citet{Corbelli:09} had chosen not to include any uniform extinction
  corrections due to the presence of a gradient of B-R index (E.~Corbelli,
  private communication).} \citet{Corbelli:09} fit the values of
$\Upsilon_\text{disk}$ rather than fixing it to a theoretically preferred
value.  The mass-to-light ratio of the disk of a spiral galaxy is known to be
poorly constrained in such a procedure since the contributions of the disk and
DM halo are similar (see e.g.  discussions in~\citet{Widrow:05,Chemin:09}).
\citet{Corbelli:09} analyzed variety of DM models and mass-to-light ratios,
providing good fit to the data.  The mass model that maximizes the
contribution of the stellar disk ($ \Upsilon_\text{disk} = 8$) has the core of
the Burkert profile \citep{Burkert:95} $r_B = 28\kpc$ (if one imposes
additional constraint on the total mass of M31 within several hundred kpc).
The DM column density remains in this model remains essentially flat from the
distance $\sim r_B$ inward (see Fig.\ref{fig:DM_distrib-m31-2010}).  Notice,
that the ``maximum disk'' fitting of~\citet{Kerins:03} (blue dashed
double-dotted line on Fig.\ref{fig:DM_distrib-m31-2010}) has higher DM
content. Therefore in this work we will adopt the Burkert DM model
of~\citet{Corbelli:09} as a \textit{minimal} possible amount of dark matter,
consistent with the rotation curve data on M31. The corresponding DM column
density is factor 2--3 lower than the one given by the previously adapted
model M31B in the inner $10$~kpc.  In what follows we will provide the
restrictions for both M31B (as the \textit{most conservative} among physically
motivated models of DM distribution in M31) and Burkert model
of~\citet{Corbelli:09}.

\subsection{M31 central part} 
\label{sec:M31_on}

Three observations of the central part of M31 (obsIDs \texttt{0112570401},
\texttt{0109270101} and \texttt{0112570101}) were used in B08 to search for a
dark matter decay signal.  After the removal of bright point sources, the
diffuse spectrum was extracted from a ring with inner and outer radii $5'$ and
$13'$, centered on M31\footnote{We adopt the distance to M31 $D_L =
  784\kpc$~\citep{Stanek:98} (at this distance $1'$ corresponds to
  $0.23\kpc$).} (see B08 for details, where this region was referred to as
\texttt{ring5-13}).  We call these three observations collectively
\texttt{M31on} in what follows. The baseline \texttt{powerlaw} model has the
total $\chi^2$ 1328 for 1363 d.o.f. (reduced $\chi^2 = 0.974$).

\begin{table} \centering \begin{tabular}[c]{|c|c|c|}
    \hline ObsID & Cleaned exposure [ks] & FoV
    [$\mathrm{arcmin}^2$] \\
    & (MOS1 / MOS2 / PN) & (MOS1 / MOS2 / PN)\\
    \hline
    \texttt{0109270101} & 16.8 / 16.7 / 15.3  & 335.4/ 336.0 / 283.6\\
    \hline
    \texttt{0112570101} & 39.8 / 40.0 / 36.0  & 332.9/ 333.1 / 285.9\\
    \hline
    \texttt{0112570401} & 29.8 / 29.9 / 23.5  & 335.6/ 336.1 / 289.5\\
    \hline \end{tabular} \caption{Cleaned exposures and FoV after the removal
    of point sources and OOT events (calculated using BACKSCAL keyword) of
    three \texttt{M31on} observations. } \label{tab:m31on} \end{table} 

Let us estimate the improvement of the signal-to-noise ratio~(\ref{eq:3}),
assuming the DM origin of LK09 spectral feature.  The average DM column
density in the model M31B of~\citet{Widrow:05}) is $ \S_{\rm M31on} =
606\:M_\odot \pc^{-2} $.  The dark matter column density from the Milky Way
halo in this direction is $74.1\:M_\odot \pc^{-2}$.

\begin{figure*}
  \includegraphics[width=\textwidth,angle=0]{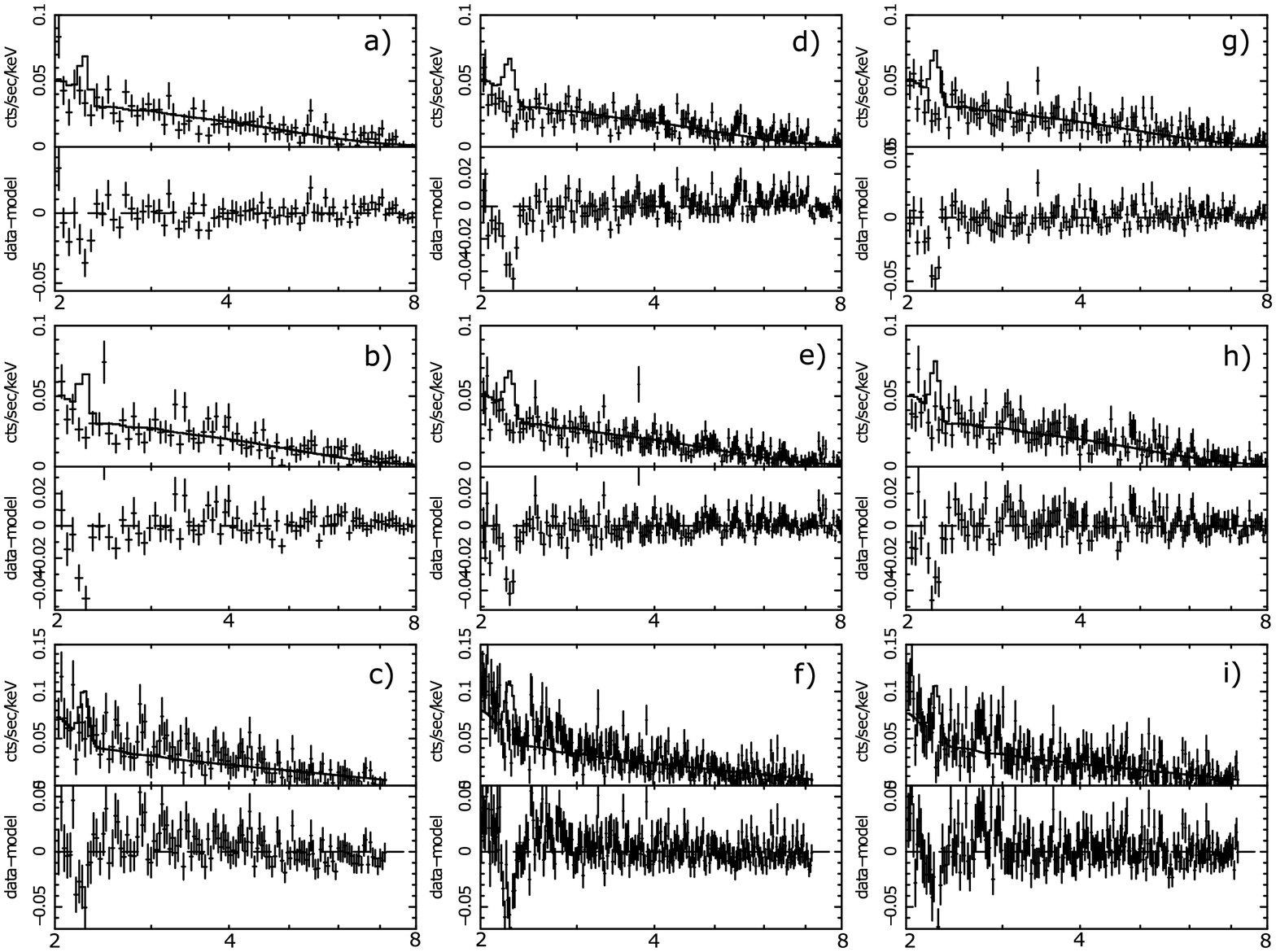} \caption{Spectra of
    three \xmm\ observations of M31 (\texttt{M31on}, Sec.~\ref{sec:M31_on}).
    The Gaussian line, obtained by proper scaling of the result of LK09 is
    also shown. The top row is for MOS1, the middle row -- for MOS2, and the
    bottom row for PN cameras.  The spectra in the left column are for the
    observation \texttt{0109270101}, the middle column -- \texttt{0112570101}
    and the right column -- for \texttt{0112570401}. The error bars include
    $5\%$ of the model flux as an additional systematic error.  For all these
    spectra combined together, the $\chi^2$ increases by \emph{at least}
    $22.2^2$, when adding a narrow Gaussian line in any position between
    $2.3\kev$ and $2.72\kev$ (see text).}  \label{fig:m31on}
\end{figure*}

The ratio of $t_\text{exp}
\times \Omega_{\fov}\times \text A_\text{eff}$ of all observations of
\texttt{M31on} (Table~\ref{tab:m31on}) and the LK09 observation is $12.9$.
Thus one expects the following improvement of the $S/N$ ratio:
\begin{equation} \label{eq:1} \frac{(S/N)_{\rm M31on}}{(S/N)_{\rm W1}} =
  \frac{606 + 74.1}{208.5 + 73.9}\sqrt{12.90} \approx 8.65\;,
\end{equation} i.e. the $\sim 2.5\sigma$ signal of LK09 should become a
prominent feature (formally about $\sim 21.6\sigma$ above the background) for
the \texttt{M31on} observations.  As described in
Section~\ref{sec:data-analysis} we add to the baseline powerlaw spectrum a
narrow Gaussian line with the normalization fixed at
$F_{\rm M31on} = \frac{606 + 74.1}{208.5 + 73.9} F_{\rm W1} \approx 1.08
\times 10^{-7}\:\mathrm{photons~ cm^{-2}~s^{-1}~ arcmin^{-2}}$. 
The quality of fit becomes significantly worse (see Fig.~\ref{fig:m31on}).
The increase of the total $\chi^2$ due to the adding of a line is equal to
$(23.2)^2$, $(22.9)^2$ and $(22.2)^2$ for $1\sigma$, $2\sigma$ and $3\sigma$
intervals with respect to the central  respectively.

\begin{figure*}
    \includegraphics[width=14cm]{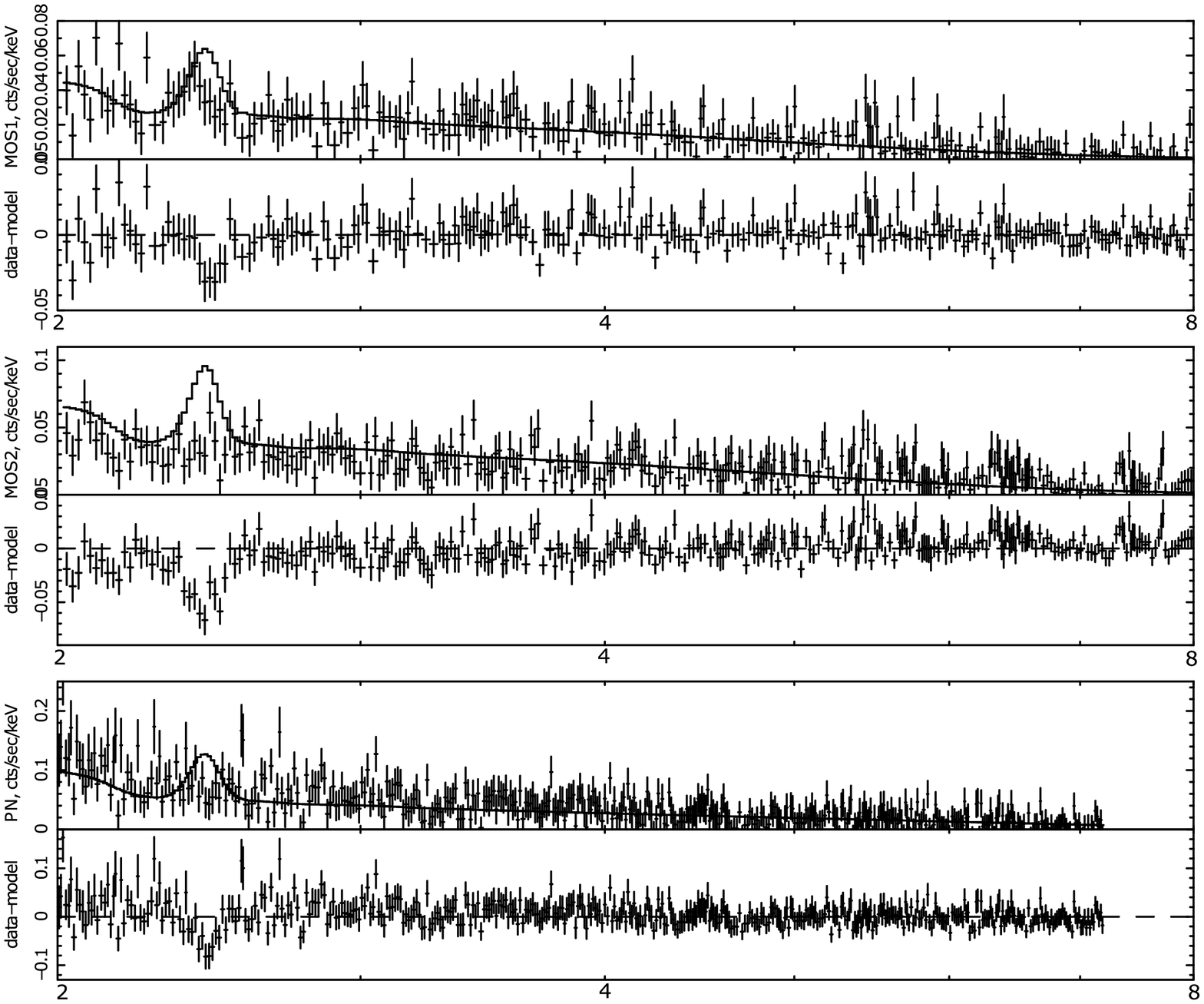} \caption{The spectra of
      off-center M31 observation \texttt{0402560301} (\texttt{M31off},
      Sec.\ref{sec:m31-outskirts}). The decaying dark matter signal, obtained
      by proper scaling of the result of~LK09 is also shown.  The error bars
      include $5\%$ of the model flux as an additional systematic error.
      Fitting these spectra together excludes the properly scaled line
      of~LK09, at the level of at least $10.4\sigma$ (see text). }
    \label{fig:m31off}
\end{figure*}

In addition to the \texttt{M31on} observations, we processed an \xmm
observation \texttt{0402560301}, positioned $\approx 22'$ off-center M31 (RA =
00h40m47.64s, DEC = +41d18m46.3s) -- \texttt{M31off} observation, see
Table~\ref{tab:m31off}.  We collected the spectra from the central $13'$
circle. Several point sources were manually excluded from the source spectra.
The rest of data reduction is described in the Section~\ref{sec:M31_on}. The
fit by the \texttt{powerlaw} model is excellent, the total $\chi^2$ equals to
895 for 997 d.o.f. (the reduced $\chi^2 = 0.898$).  Using the DM estimate
based on the model M31B we find that the average DM column density for the
\texttt{M31off} $\S_{\rm M31off} = 388.6 M_\odot\pc^{-2}$ plus the Milky Way
halo contribution $74.3 M_\odot\pc^{-2}$.  The estimate of the line
significance is similar to the previous Section and gives 
\begin{equation}
  \label{eq:4} \frac{(S/N)_{\rm M31off}}{(S/N)_{\rm W1}} = \frac{388.6 +
    74.3}{208.5 + 73.9}\sqrt{8.76} \approx 4.85 
\end{equation} 
and therefore, under the assumption of DM nature of the feature of LK09, one
would expect $\sim 12.1\sigma$ detection.  After that, we add a narrow line
with the normalization $F_{\rm M31off} \approx 7.37 \times
10^{-8}~\mathrm{photons~cm^{-2}~s^{-1}~arcmin^{-2}}$ and perform the
procedure, described in Section~\ref{sec:data-analysis}.  The observation
\texttt{M31off} rules out the DM decay line origin of the LK09 feature with
high significance (see Fig.~\ref{fig:m31off}): the increase of $\chi^2$ due to the addition of this line
has the minimum value $\Delta \chi^2\simeq (10.4)^2$ at 2.44~keV, which is
within $1\sigma$ interval of the quoted central value of the LK09 feature.

We also perform the analysis, adding a line, whose flux is determined
according to the DM density estimates based on the model
of~\citet{Corbelli:09}, that we consider to be a minimal DM model for M31 (as
discussed in~\ref{sec:dm-m31}).  For this model the column density in the
central $1$--$3\kpc$ decreases by a factor $\sim 3.4$ as compared with the
value, based on M31B.  \citet{Kusenko:10a} claimed that in this case the LK09
line becomes consistent with the \texttt{M31on} observations.  However, our
analysis shows that the total $\chi^2$ \emph{increases}, when adding the
corresponding line to the model, by $(5.7)^2$ for $3\sigma$ variation of the
position of the line. The corresponding increase of total $\chi^2$ for
\texttt{M31off} region is $(2.0)^2$.  Combining \texttt{M31on} and
\texttt{M31off} observations, one obtains $(6.2)^2$ increase of the total
$\chi^2$.

We conclude therefore that despite the uncertainties in DM modeling, the
analysis of diffuse emission from the central part of M31 ($1$--$8\kpc$ off
the center), as measured by \xmm, disfavors the hypothesis that the spectral
feature, observed in Willman~1, is due to decaying dark matter. Nevertheless,
to strengthen this conclusion further we analyzed available \xmm\ observations
of M31 in the region $10$--$20\kpc$ off-center, where the uncertainties in the
mass modeling of M31 reduce significantly as compared with the central
$5-8\kpc$~(c.f.~\citealt{Chemin:09,Corbelli:09}, see also
Section~\ref{sec:dm-m31}, in particular Fig.~\ref{fig:DM_distrib-m31-2010}).

\begin{table}
  \centering \begin{tabular}[c]{|c|c|c|} \hline ObsID & Cleaned exposure
    [ks] & FoV
    [$\mathrm{arcmin}^2$] \\
    & (MOS1 / MOS2 / PN) & (MOS1 / MOS2 / PN)\\
    \hline
    \texttt{0402560301} & 41.9 / 42.2 / 35.2  & 405.6/ 495.4 / 433.3\\
    \hline \end{tabular} \caption{Cleaned exposures and FoV (calculated using
    BACKSCAL keyword) of the observation \texttt{M31off} (obsID
    \texttt{0402560301}). The significant difference in FoVs between MOS1 and
    MOS2 cameras is due to the loss CCD6 in MOS1 camera.  } \label{tab:m31off}
\end{table} 

\subsection{M31 off-center (10-20 kpc)}
 \label{sec:m31-outskirts} 

\begin{table}
\centering 
\begin{tabular}[c]{|c|c|c|} \hline 
ObsID & Cleaned exposure [ks] & FoV [$\mathrm{arcmin}^2$] \\
& (MOS1 / MOS2 / PN) & (MOS1 / MOS2 / PN)\\ \hline
0511380101 & 44.3 / 44.5 / 37.6 & 356.2 / 400.4 / 333.3\\ \hline
0109270401 & 38.5 / 38.4 / 33.5 & 387.8 / 384.1 / 366.7\\ \hline
0505760401 & 26.0 / 26.0 / 21.7 & 330.5 / 374.1 / 325.0\\ \hline
0505760501 & 23.8 / 23.8 / 19.8 & 344.8 / 395.2 / 313.2\\ \hline
0402561301 & 22.7 / 22.7 / 20.1 & 367.2 / 419.5 / 382.6\\ \hline
0402561401 & 39.3 / 39.3 / 33.7 & 349.9 / 408.3 / 343.6\\ \hline
0402560801 & 42.6 / 42.6 / 36.6 & 373.6 / 435.8 / 393.7\\ \hline
0402561501 & 38.5 / 38.5 / 34.0 & 369.3 / 421.0 / 407.5\\ \hline
0109270301 & 24.4 / 24.6 / 22.2 & 443.4 / 439.6 / 405.0\\ \hline
\end{tabular} 
\caption{Cleaned exposures and FoV after the removal of point sources and OOT
  events (calculated using BACKSCAL keyword) of nine M31 observations off-set
  by more than 10~kpc from the center (\texttt{M31out} observations).} 
\label{tab:m31out}
\end{table}

 We selected 9 observations with MOS cleaned exposure
greater than 20~ks (Table~\ref{tab:m31out}).  The total exposure of these
observations is about 300~ks. Based on the statistics of these observations
one would expect a detection of the signal of LK09 with the significance
$12.1\sigma$ (for M31B) and $11.2\sigma$ (for~\citet{Corbelli:09}).  The fit
to the baseline model is very good, giving total $\chi^2$ equal to 5931 for
5883 d.o.f. (the reduced $\chi^2 = 1.008$).  However, adding the properly
rescaled line significantly reduced the fit quality, increasing the total
$\chi^2$ by $(12.0)^2$/$(10.7)^2$/$(10.7)^2$ when varying line position within
$1\sigma$, $2\sigma$ and $3\sigma$ intervals (using M31B model) and by
$(11.7)^2$, $(10.7)^2$ and $(10.6)^2$ for $1\sigma$, $2\sigma$ and $3\sigma$
intervals (using the model of~\citealt{Corbelli:09}, which gives DM column density in \texttt{M31out} about $160-180 M_\odot \pc^{-2}$ (including Milky Way contribution), see Fig.~\ref{fig:DM_distrib-m31-2010}).

\subsection{Combined exclusion from M31}
\label{sec:comb-excl-from}

Finally, performing a combined fit to all 13 observations of M31
(\texttt{M31}, \texttt{M31off}, \texttt{M31out}) we obtain the exclusion of
more than $26\sigma$ (using the DM model M31B of~\cite{Widrow:05}) and more
than $13\sigma$ for the minimal DM model of~\citet{Corbelli:09}. As described
in Section~\ref{sec:data-analysis}, in deriving these results we allowed the
normalization of the baseline \texttt{powerlaw} model to vary independently
for each camera, observing the same spatial region, added additional 5\% of
the model flux as a systematic uncertainty and allowed the position of the
narrow line to vary within $2.3 - 2.72\kev$ interval.

\section{Dwarf spheroidal galaxies}
\label{sec:dwarf-spher-galax} 

Since Willman~1 is purported to be a dwarf spheroidal (dSph) galaxy (but see section~\ref{subsubsec:willman1}), as the next step in examining the hypothesis of LK09 we compare
estimates of the dark mass of Willman~1 to dark masses estimated for other
dSph satellites of the Milky Way.  Specifically we consider two of the
(optically) brightest dSphs, Fornax and Sculptor, for which comparable
X-ray data exist.  To characterize the dark matter halos of dSphs, we
adopt the general DM halo model of~\cite{Walker:09}, with density profile given~by 
\begin{equation}
  \rho(r)=\rho_s\biggl (\frac{r}{r_s}\biggr )^{-\gamma}\biggl [1+\biggl
  (\frac{r}{r_s}\biggr )^{\alpha}\biggr ]^{\frac{\gamma-3}{\alpha}},
  \label{eq:hernquist1} \end{equation} where the parameter $\alpha$ controls
the sharpness of the transition from inner slope $\lim_{r\rightarrow
  0}d\ln(\rho)/d\ln(r)\propto -\gamma$ to outer slope $\lim_{r\rightarrow
  \infty}d\ln(\rho)/d\ln(r)\propto -3$.  This model includes as special cases
both \emph{cored} ($\alpha=1, \gamma=0$) and NFW~\citep{Navarro:96}
($\alpha=\gamma=1$) profiles.  Assuming that dSph stars are spherically
distributed as massless test particles tracing the DM potential, the DM
density relates to observables---the projected stellar density and velocity
dispersion profiles---via the Jeans Equation (see Equation 3 of
\citet{Walker:09}).  \citet{Walker:09} show that while the data for a given
dSph do not uniquely specify any of the halo parameters individually, the bulk
mass enclosed within the optical radius is generally well constrained, subject
to the validity of the assumptions of spherical symmetry and dynamic
equilibrium and negligible contamination of stellar velocity
samples from unresolved binary-orbital motions.  

\subsection{Willman~1} 
\label{subsubsec:willman1} 

\begin{figure*} \centering
  \includegraphics[width=\textwidth]{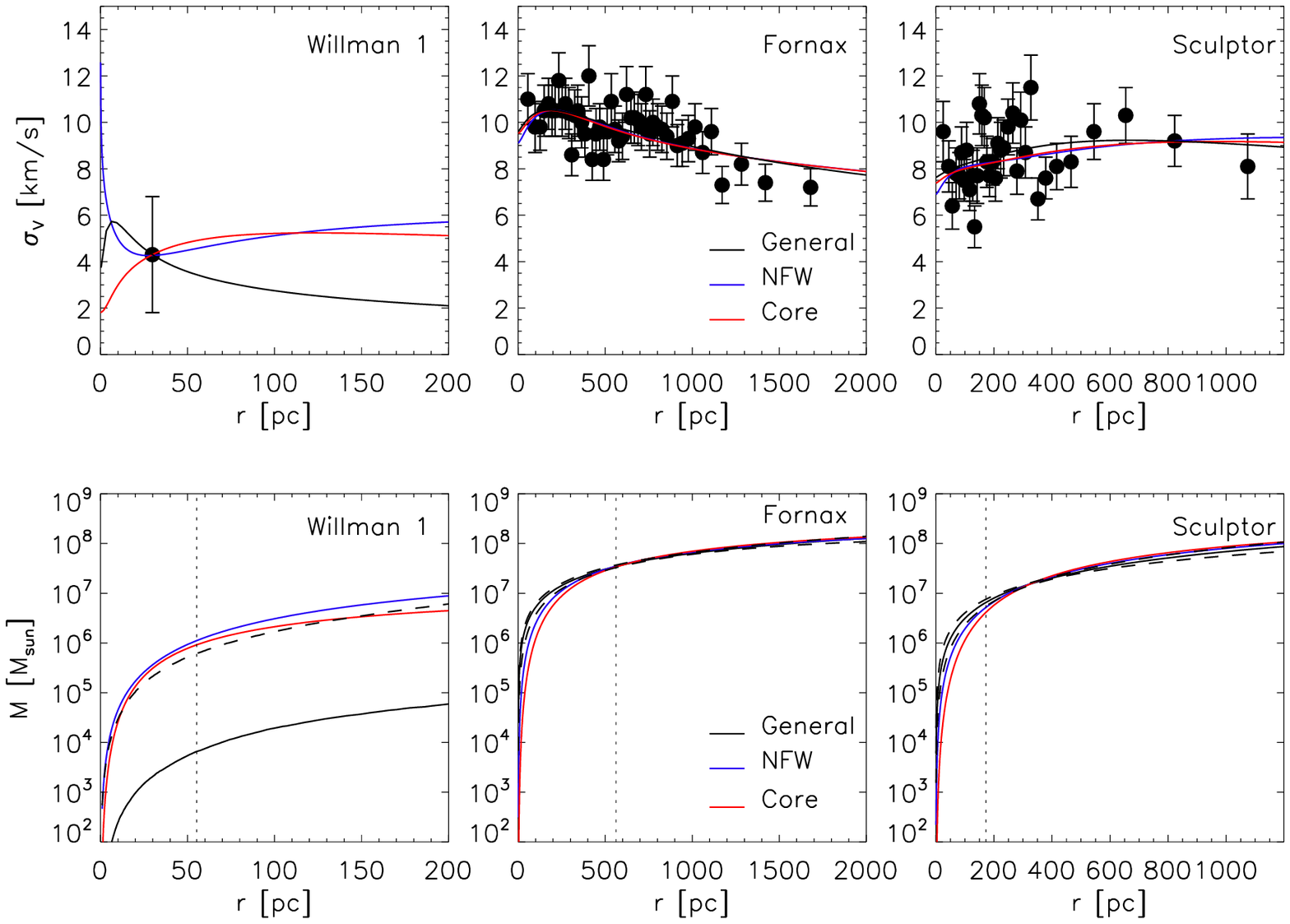} \caption{DM mass
    modeling in Fornax and Willman~1.  Top panels display empirical velocity
    dispersion profiles as calculated by \protect\cite{Walker:09} for Fornax,
    Sculptor and from the Willman~1 data of \protect\cite{Martin:07} (for
    Willman~1, the small published sample of 14 stars allows for only a single
    bin).  Overplotted are the median profiles obtained from the MCMC analysis
    described by \protect\cite{Walker:09}, using the general dark matter halo
    model~(\ref{eq:hernquist1}), as well as the best-fitting NFW and cored
    halo models.  Bottom panels indicate the corresponding
    spherically-enclosed mass profiles.  In the bottom panels, dashed lines
    enclose the central $68\%$ of accepted general models (for Willman~1, the
    lower bound falls outside the plotting window).  Vertical dotted lines
    indicate the outer radius of the field of view of the X-ray observations
    ($14'$ for Fornax, $7.5'$ for Sculptor and $5'$ for Willman~1). }
  \label{fig:forwil} \end{figure*}

Despite some indications that Willman~1 is a dark-matter dominated dSph galaxy
\citep{Martin:07,Strigari:07a}, it should be noted that the galactic status of
Willman~1 remains uncertain.  Claims that Willman~1 is a \emph{bona fide}
galaxy---as opposed to a star cluster devoid of dark matter---stem primarily
from two considerations.  First, under simple dynamical models, the stellar
velocity dispersion ($4.3\pm 2.5$ km s$^{-1}$; \citet{Martin:07}) of Willman~1
implies a large mass-to-light ratio ($M/L_V\sim 700 M_{\odot}/L_{V,\odot}$),
indicative of a dominant dark-matter component like those that characterize
other dSph galaxies.  Second, the initial spectroscopic study by
\citet{Martin:07} suggests that Willman~1 stars have a metallicity range $-2.0
\leq \mathrm{[Fe/H]} \leq -1.0$, significantly broader than is observed in
typical star clusters.  However, both arguments are vulnerable to scrutiny.
For example, if the measured velocity dispersion of Willman~1 receives a
significant contribution from unresolved binary stars and/or tidal heating,
then the inferred mass may be significantly overestimated.  Demonstrations
that binary orbital motions contribute negligibly to dSph velocity dispersions
have thus far been limited to intrinsically ``hotter'' systems, with velocity
dispersions $\sigma_V\sim 10$ km s$^{-1}$ \citep{Olszewski:95,Hargreaves:96}.
Furthermore, given Willman~1's low luminosity, kinematic samples must include
faint stars close to the main-sequence turnoff (see Figure 8 of
\cite{Martin:07}).  These stars are physically smaller than the bright red
giants observed in brighter dSphs, and thus admit tighter, faster binary
orbits.  Therefore, the degree to which binary motions may inflate the
estimated mass of Willman~1 remains unclear.  Finally, follow-up spectroscopy
by \cite{Siegel:08} suggests that the relatively high-metallicity stars
observed by \cite{Martin:07} are in fact foreground contaminants contributed
by the Milky Way along the line of sight to Willman~1.  Removal of these stars
from the sample would erase the large metallicity spread reported by
\cite{Martin:08}, and the remaining stars would have a narrow metallicity
distribution consistent with that of a \emph{metal-poor globular
  cluster}~\citep{Siegel:08}.  Nevertheless, we proceed under the assumption
that Willman~1 is indeed a dSph galaxy for which the measured stellar velocity
dispersion provides a clean estimate of the dark matter content.  This
assumption gives conservative estimates of the expected DM decay signal in
other objects (including M31), since any overestimate of the DM column density
in Willman~1 will result in an underestimate of the relative S/N expected in
other objects (see Eq.~(\ref{eq:3})).  For Willman~1 we adopt the
line-of-sight velocity data for 14 member stars observed by \citet{Martin:07}.
These stars, which lie a mean distance of $30\pc$ in projection from the
center of Willman~1, have a velocity dispersion of $\sigma_V=4.3\pm 2.5$ km
s$^{-1}$.  To specify the stellar surface density, we adopt a Plummer profile,
$I(R)=I_0[1+R^2/r_{\rm half}^2]^{-2}$, with parameters $r_{\rm half}=25 \pc$
and $I_0=0.5 L_{\odot}\pc^{-2}$ as measured by \cite{Martin:08}.  Of course
the single point in the empirical velocity dispersion ``profile'' relates no
information about the \emph{shape} of the profile, and these data therefore
provide only weak constraints on the dark matter content of Willman~1.  The
solid black line in the bottom-left panel of Figure \ref{fig:forwil} displays
the median spherically-enclosed mass profile for the general halo model of
Equation \ref{eq:hernquist1}, with dotted lines enclosing the region
corresponding to the central $68\%$ of posterior parameter space from the
Markov-Chain Monte Carlo analysis (see \cite{Walker:09} for a detailed
description of the MCMC procedure).  The small published kinematic data set
(14 stars) for Willman~1 allows for only a single bin in the velocity
dispersion profile (see the upper right panel in Fig.~\ref{fig:forwil} and
compare it e.g. with Fornax dSph (the left panel)).  This resulting
constraints on the mass profile of Willman~1 are therefore very poor.  The
allowed masses \emph{span several orders of magnitude at all radii of
  interest} and are consistent with negligible dark matter content---the lower
bound ($68\%$~confidence interval) on the mass within $200\pc$ (including all
of the Willman~1) is as low as $100 M_{\odot}$.  \citet{Strigari:07a} obtained
for Willman~1 $M(<100\pc) = 1.3^{+1.5}_{-0.8}\times 10^6M_\odot$ with the DM
column density for the best-fit values of parameters being $\S_{\rm W1} \sim
200 M_\odot\pc^{-2}$.  The discrepancy with results of \citet{Strigari:07a}
may arise from two possible sources.  First, \cite{Strigari:07a} use an
independent, still unpublished kinematic data set (B. Willman et al., in
preparation) for Willman~1.  Although the global velocity dispersion is
statistically equivalent to that which we measure from \citet{Martin:08}
sample, the data set used by \cite{Strigari:07a} is larger, containing 47
Willman~1 members.  Second, \cite{Strigari:07a} adopt more stringent priors
motivated by cosmological N-body simulations, under which $\rho_s$ and $r_s$
are strongly correlated \citep{Bullock:07,Diemand:07}.  We have used more
general models here in order to help illustrate the uncertainties in the mass
modeling of Willman~1.  However, in the interest of obtaining the most
conservative estimates for the signal we expect to see from other objects if
the LK09 detection is real, we shall continue to adopt one of the largest
available estimates of the dark matter column density in Willman~1---namely,
the value of $\S_{\rm W1}=208.5 M_{\odot}\pc^{-2}$ that results from the
modeling of \cite{Strigari:07a}.   

\subsection{Fornax and Sculptor dSphs}
\label{sec:fornax} 

\begin{table} \centering \begin{tabular}[c]{lcccc}
    &\multicolumn{2}{c}{Fornax} & \multicolumn{2}{c}{Sculptor}\\
    &\scriptsize $\rho_s$ [$M_\odot \pc^{-3}$] & $r_s$
    &\scriptsize $\rho_s$ [$M_\odot \pc^{-3}$] & $r_s$\\
    \hline
    NFW  & 0.1  & 500 & 0.04 & 920\\
    Core & 0.57 & 280 & 0.48 & 350 \\
    \hline \end{tabular} \caption{Best-fit NFW ($\alpha = \gamma = 1$ in
    Eq.(\ref{eq:hernquist1})) and core profiles ($\alpha=1$, $\gamma = 0$) for
    Fornax and Sculptor dSphs} \label{tab:forsc}
\end{table} 

For Fornax and Sculptor, two of the brightest and most well-studied dSph
satellites of the Milky Way, the availability of large kinematic data sets of
$\sim 2600$ and $\sim 1400$ members, respectively~\citep{Walker:09b}, allows
us to place relatively tight constraints on the dark-matter column densities.
Figure \ref{fig:forwil} displays the empirical velocity dispersion profiles,
as well as the fits we obtain from the general (Eq. \ref{eq:hernquist1}), NFW
and cored halo models.  Adopting a Fornax distance of $\sim 138$
kpc~\citep{Mateo:98}, we integrate the projected density profile out to a
radius of $560 \pc$, which corresponds to the $14'$ region (extraction region
of \xmm\ observation, see below) and is similar to the half-light radius of
Fornax.  For Fornax, the best-fitting density profile parameters are given in
Table~\ref{tab:forsc}.\footnote{Note that the fitting parameters $\rho_s$ and
  $r_s$ are typically degenerate in the mass modeling of
  dSphs~\citep{Strigari:06,Penarrubia:07}.  Thus neither parameter is
  constrained uniquely.  The best fits considered here generally represent a
  ``family'' of models that follow a $\rho_s$, $r_s$ relation consistent with
  the data, but which all tend to have the same bulk mass enclosed within the
  optical radius \citep{Penarrubia:08}.} For NFW density profile ($V_{max} =
17.0\mathrm{km\:s^{-1}}$) gives $\S_{\for,\rm NFW} = 55.2 M_\odot\pc^{-2}$ and
the best-fitting cored profile ($V_{max} = 17.4\;\mathrm{km s^{-1}}$) gives a
slightly lower value $\S_{\for}= 54.4 M_{\odot}\pc^{-2}$ that we adopt for
subsequent analysis.

Adopting a Sculptor distance of 79 kpc \citep{Mateo:98}, we integrate the
projected density profile over a square of $175^2\pc^2$, which corresponds to
the $7.6'^{2}$ region (extraction region of \Chandra\ observations on the
ACIS-S3 chip, see below).  The best-fit parameters for Sculptor are given in
Table~\ref{tab:forsc}. The best-fitting cored profile ($V_{max} =
19.8\;\mathrm{km s^{-1}}$) gives a DM column density $\S_{\scl, \rm core}=
147M_{\odot}\pc^{-2}$ and the best-fitting NFW profile ($V_{max} =19.4
\mathrm{km\:s^{-1}}$), gives $\S_{\scl,\rm NFW} = 140.3 M_\odot\pc^{-2}$ and
that we adopt for subsequent analysis.  Notice that using the parameters of
the ``universal DM density profile'' of~\cite{Walker:09}, one would get only
slightly higher values.  In any case, the DM column densities of Fornax and
Sculptor are much more tightly constrained than that of Willman~1.  The Fornax
column density is smaller than the expected column density of Milky Way halo
in the direction of Fornax, which we estimate to be $83.2 M_\odot\pc^{-2}$.
The estimated DM column density of the Milky Way in the direction of Sculptor
is $95.7 M_\odot \pc^{-2}$ (see Section~\ref{sec:choice-observ-targ}).

 \subsection{\xmm observation of Fornax dSph}
\label{sec:xmm-forn}

Fornax is one of the most deeply observed dSphs in X-ray wavelengths. We have
processed the 100~ks \xmm observation (ObsID \texttt{0302500101}) of the
central part of Fornax dSph.  The data analysis is described in the 
Section~\ref{sec:data-analysis}. In each camera, we extract signal from the
$14'$ circle, centered on the Fornax dSph. The cleaned exposure and corresponding
fields of view (calculated using the BACKSCAL keyword) for all three cameras
(after subtraction of CCD gaps, OOT strings in PN camera and CCD~6 in MOS1
camera) are shown in Table~\ref{tab:fornax}.  Based on the DM estimates,
presented in the Section~\ref{sec:fornax} we expect the following improvement
in S/N ratio
\begin{equation}
  \label{eq:fornax}
  \frac{(S/N)_{\for}}{(S/N)_{\rm W1}} =
  \frac{54.4 + 83.2}{208.5 + 73.9}\sqrt{12.12} \approx 1.70
\end{equation}
i.e. we expect a $\sim 4.2\sigma$ signal from Fornax, assuming DM decay line
origin of the LK09 feature.

The fit to the baseline model is very good, $\chi^2 = 955$ for 983 d.o.f.
(reduced $\chi^2 = 0.972$).  Note that the parameters are consistent with
those of extragalactic diffuse X-ray background as it should be for the dwarf
spheroidal galaxy where we do not expect any diffuse emission at energies
above $2\kev$~\citep[see
also][]{Boyarsky:06c,Boyarsky:06d,Jeltema:08,Loewenstein:08,Riemer:09a}.

\begin{figure*}
  \includegraphics[width=\textwidth]{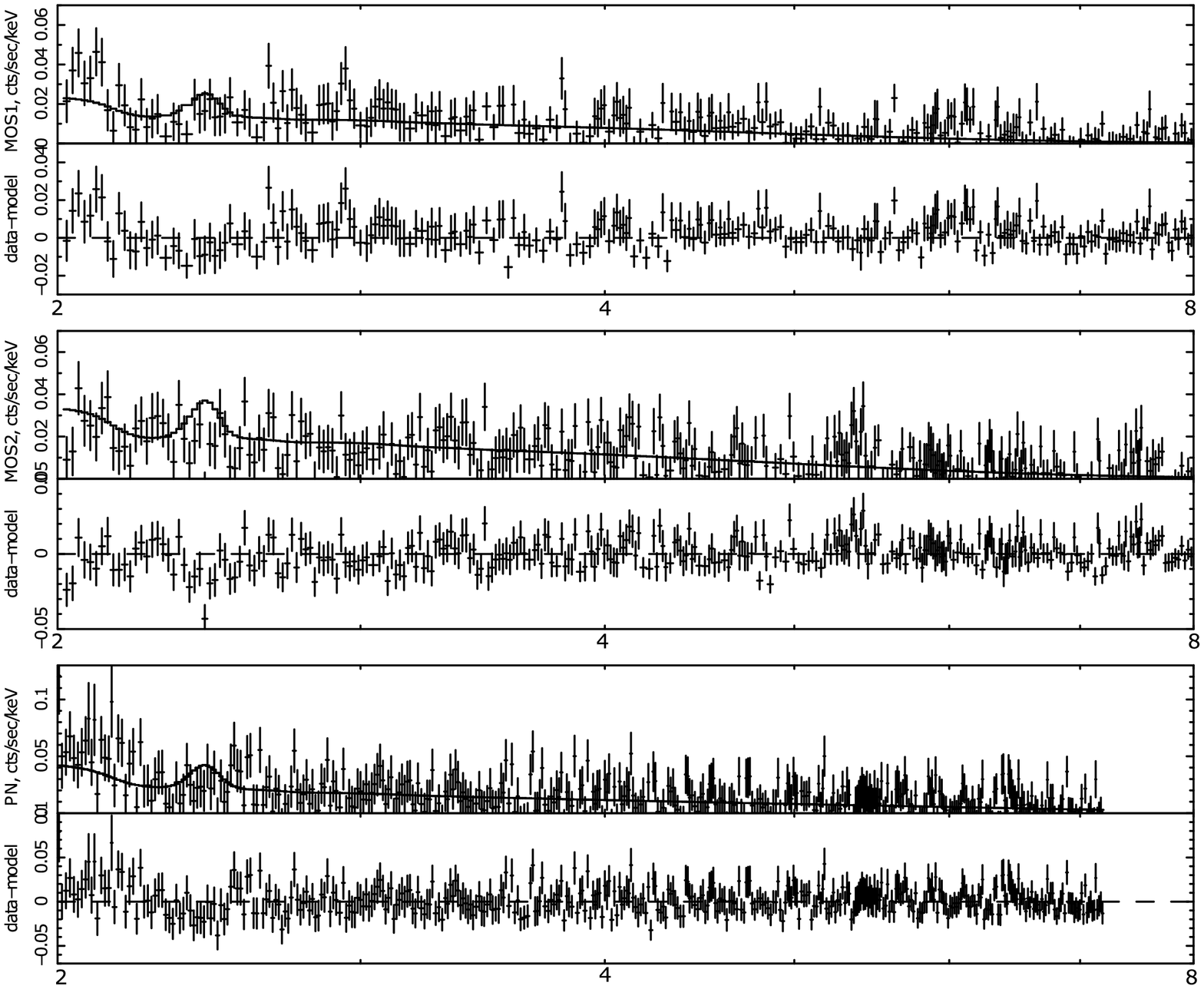}
  \caption{The spectra of the \xmm observation \texttt{0302500101},
    Sec.~\ref{sec:fornax}. The error bars include $5\%$ of the model flux as
    an additional systematic error.  Fitting these spectra together excludes
    the properly scaled LK09 line at the level of at least $2\sigma$
    ($3.3\sigma$ if one restricts the position of the line to the interval
    2.30 -- 2.72 keV) instead of expected improving of the quality of fit by
    about $4\sigma$ (if the line were of the DM origin).}
\label{fig:fornax_all}
\end{figure*}

After adding a narrow line with the normalization $F_{\for} \approx 2.19
\times 10^{-8}~\mathrm{photons~cm^{-2}s^{-1}arcmin^{-2}}$, the $\chi^2$
\emph{increases}, the minimal increase is equal to $(5.1)^2$, $(4.2)^2$ and
$(3.3\sigma)^2$ for a position of the line within $1\sigma$, $2\sigma$ and
$3\sigma$ intervals from LK09 feature, respectively (after adding $5\%$ of
model flux as a systematic error).  We see that adding a properly scaled
Gaussian line worsens $\chi^2$ by at least $\sim 3.3\sigma$ (instead of
expected \emph{improvement} of the quality of fit by a factor of about $16$ if
the LK09 feature were the DM decay line), see Fig.~\ref{fig:fornax_all} for details.

\begin{table}
  \centering
  \begin{tabular}[c]{|c|c|c|}
    \hline
    \xmm & Cleaned exposure [ks]  & FoV
    [$\mathrm{arcmin}^2$] \\
    ObsID  & (MOS1 / MOS2 / PN) & (MOS1 / MOS2 / PN)\\
    \hline
    \texttt{0302500101} & 53.8 / 53.9 / 48.2  & 459.1 / 548.5 / 424.9\\
    \hline
  \end{tabular}
  \caption{Cleaned exposures of the \xmm\ observation of Fornax dSph}
  \label{tab:fornax}
\end{table}

The combination of all \xmm observations used in this work (\texttt{M31on},
\texttt{M31off}, \texttt{M31out} and \texttt{Fornax}) provide a minimum of
$26.7\sigma$ exclusion at 2.35~keV (which falls into the $3\sigma$ energy
range for the LK09 spectral features), after adding a 5\% systematic error.
Within $1\sigma$ and $2\sigma$ energy intervals, the minimal increase of
$\chi^2$ is $29.0^2$ and $27.3^2$, correspondingly.  Therefore, one can
exclude the DM origin of the LK09 spectral feature by $26\sigma$. This
exclusion is produced by using the most conservative DM model M31B
of~\cite{Widrow:05}.  Using the minimal DM model of~\cite{Corbelli:09}, one
can exclude the LK09 feature at the level more than $14\sigma$.

\subsection{{\it Chandra} observations of the Sculptor dSph}
\label{sec:Sculptor}

\subsubsection{The data and data preparation}

For the Sculptor dSph there is one observation of $50\ks$ (ObsId
\texttt{9555}) and 21 observations of $5-6\ks$ (ObsID \texttt{4698-4718}), see Table~\ref{tab:obs} for details. In all the observations, the centre of Sculptor is on the ACIS-S3 chip, and we
have restricted the analysis to this chip.

Throughout the entire analysis we used \texttt{CIAO 4.1} with \texttt{CALDB 4.1}
\citep{CIAO_general} and \texttt{XSPEC 12.4} \citep{xspec}. All the data were observed
using the VFAINT telemetry mode and thus required reprocessing prior to any
data analysis in order to take advantage of the improved background event
rejection based on $5\times5$ pixel islands instead of the normal $3\times3$
pixel.

We used the wavelet algorithm {\it wavdetect} in \texttt{CIAO} to find and exclude
point sources for each observation. The removed areas are very small compared
to the field of view, and were consequently neglected in the following
analysis. Furthermore the observations were lightcurve cleaned using {\it
  lc\_clean}. The observations \texttt{4705} and \texttt{4712} were flared and
left out in the subsequent analysis. The number of point sources removed and
the exposure time after lightcurve cleaning are given in \tabref{obs}. The
variation in the number of point sources is mainly due to slightly different
observed areas. The observations are all centered on the same spot, but there
is a variation in detector roll angles. Assuming spherical symmetry of
Sculptor, this has no importance for the further analysis.

\begin{table} 
  \begin{tabular}{l|l|r|r}
  ObsID & Obs date & Exposure$^a$ & Point sources$^b$ \\ \hline
  \texttt{4698}		& 2004 Apr 26 		& 6060~s		& 9 \\
  \texttt{4699}		& 2004 May 7		& 6208~s		& 6 \\
  \texttt{4700}		& 2004 May 17		& 6104~s		& 8 \\
  \texttt{4701}		& 2004 May 30		& 6069~s		& 8 \\
  \texttt{4702}			& 2004 Jun 12		& 5883~s		& 7 \\
  \texttt{4703}			& 2004 Jun 27		& 5880~s		& 8 \\
  \texttt{4704}			& 2004 Jul 12		& 5915~s		& 9 \\
  \texttt{4705}			& 2004 Jul 24		& ---			& 6 \\
  \texttt{4706}			& 2004 Aug 4		& 6075~s		& 8 \\
  \texttt{4707}			& 2004 Aug 17		& 5883~s		& 8 \\
  \texttt{4708}			& 2004 Aug 31		& 5880~s		& 8 \\
  \texttt{4709}			& 2004 Sep 16		& 6091~s		& 10 \\
  \texttt{4710}			& 2004 Oct 1		& 5883~s		& 8 \\
  \texttt{4711}			& 2004 Oct 11		& 5727~s		& 9 \\
  \texttt{4712}			& 2004 Oct 24		& ---			& 7 \\
  \texttt{4713}			& 2004 Nov 5		& 6073~s		& 11 \\
  \texttt{4714}			& 2004 Nov 20		& 5649~s		& 13 \\
  \texttt{4715}			& 2004 Dec 5		& 5286~s		& 10 \\
  \texttt{4716}			& 2004 Dec 19		& 6016~s		& 8 \\
  \texttt{4717}			& 2004 Dec 29		& 6073~s		& 11 \\
  \texttt{4718}			& 2005 Jan 10		& 6060~s		& 10 \\ \hline
  short obs combined & ---			& 113169~s	& --- \\ \hline
  \texttt{9555}			& 2008 Sep 12		& 48935~s	& 21 \\ \hline
  All obs  & --- & 162104~s & --- \\ \hline
\end{tabular}
\caption{The analyzed \Chandra{} observations of Sculptor Dwarf. $^a$The exposure times after lightcurve cleaning using {\it lc\_clean}. $^b$The number point sources found by {\it wavdetect} and removed.} 
    \label{tab:obs}
  \end{table}

\subsubsection{The spectra}
The spectra were extracted from a square region of $7.6\arcmin\times7.6\arcmin$ covering most of the ACIS-S3 chip but excluding the edges where the calibration is less precise \citep{CIAO_analysis}. For later fitting they where binned to at least 15 counts per bin.

The spectra of the short observations (\texttt{4698-4718}, except
\texttt{4705}, \texttt{4712}) were combined into one observation, see Fig.~\ref{fig:Sculptor_spectra}. This is justified because they fulfill the following criteria: They are observations
of the same (non-variable) object and thus have similar count rates, they are
extracted from the same region of the same chip, their rmf's (response matrix
functions) are extremely similar. The spectra were added using the \texttt{FTOOL} {\it
  mathpha} \citep{ftools} with the uncertainties propagated as if they were
pure Gaussian.

\subsubsection{Background subtraction}
We subtracted only the particle background as observed with the ACIS detector
stowed\footnote{The files can be downloaded from
  http://cxc.harvard.edu/contrib/maxim/acisbg/}. Around 2005 there was a
change in the spectral shape of ACIS-S3 and consequently new particle
background files were produced \citep[][]{Markevitch:webpage}. We used the new
particle background files (2005-2009) for both spectra. The particle
background was normalised in the $10-12\keV$ interval where any continuous
emission from Sculptor is negligible \citep{Markevitch:blank}. For \texttt{9555} the
effect of changing the normalization interval to e.~g. $4-6\keV$ is less than
3\%. For the combined observations there is a larger excess at low energies,
which makes it more sensitive to the choice of normalization interval and
consequently we chose the conservative approach with the $10-12\keV$ interval.
The larger excess might come from point sources, which are unresolved in the
short exposures, but resolved and thus removed in the long exposure. The
excess is mainly at energies below $2\keV$ and are irrelevant to the present
analysis.

\begin{figure*}
\centering
\includegraphics[width=0.6\textwidth,angle=-90]{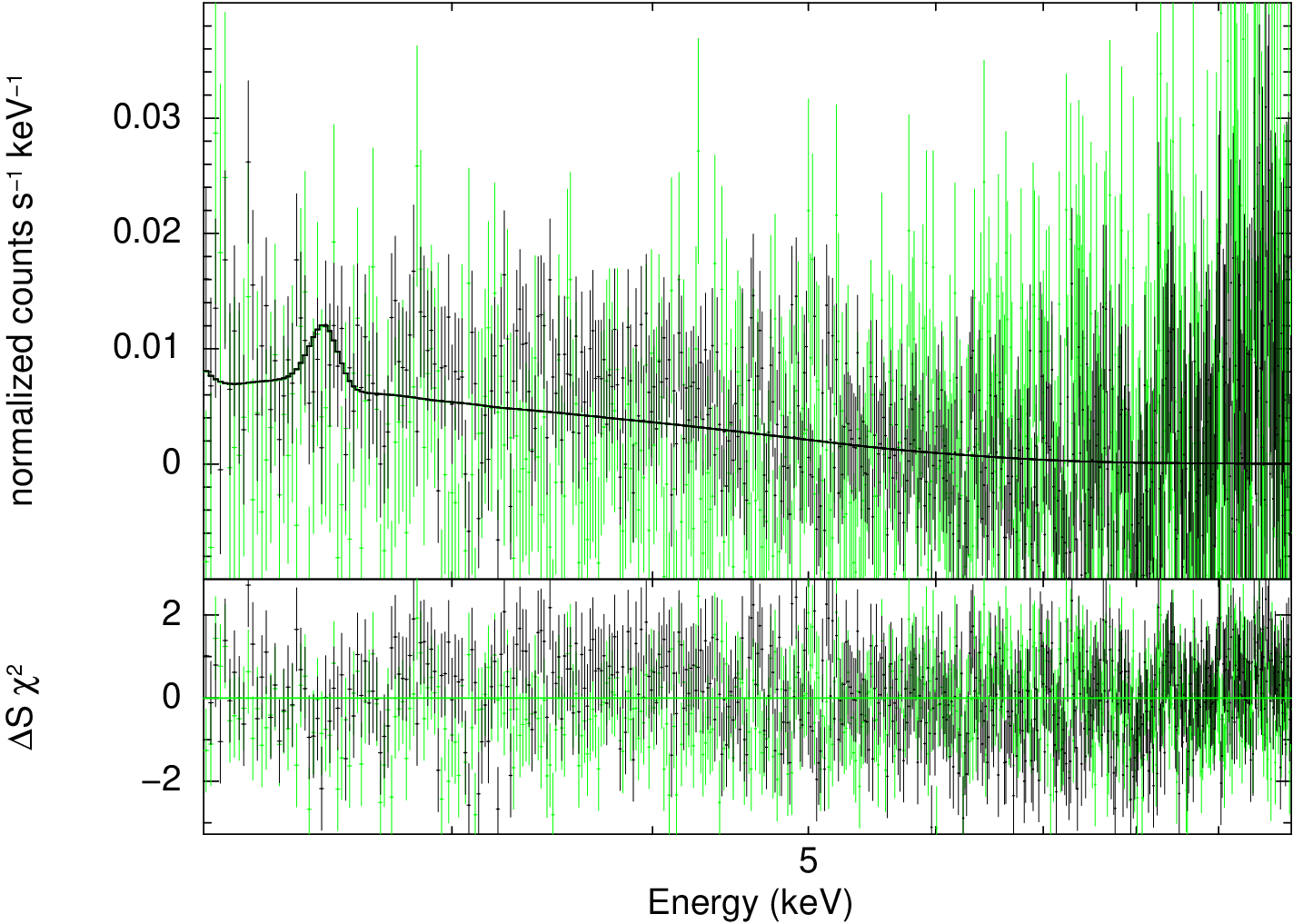}
\caption{The background subtracted \Chandra{} spectra of Sculptor dSph with
  the short combined observations in black and ObsID \texttt{9555} in a lighter (green)
  color. Also shown the best fit power law model fitted to both observations
  simultaneously (index fixed to $1.46$) and a Gaussian corresponding to the
  expected DM line emission as suggested by LK09. The dip in the
  residuals is a visual indication of the absence of any line signal at $\sim
  2.5 \keV$.}
	\label{fig:Sculptor_spectra}
\end{figure*}

The background subtracted spectra are shown in \figref{Sculptor_spectra} with
the combined observations in black and ObsID \texttt{9555} in a lighter
(green) color.  Also shown is the best fit to a power law model and the
Gaussian expected according to the signal proposed by LK09 which is further
discussed below.

\subsubsection{Constraints}

The total exposure of Sculptor is 162.1~ks which is $1.63$ times longer
than the Willman~1 observation. Sculptor is observed with the ACIS-S3 chip,
which at $2.5\keV$ has an effective area that is $1.23$ times larger than that
of ACIS-I which Willman~1 is observed with. The Sculptor spectra are extracted
from a $(7.6\arcmin)^2$ square, which is $0.74$ of the $5\arcmin$ circle for
which the Willman~1 spectra were extracted.
\begin{equation}
  \frac{(S/N)_{\rm Sc}} {(S/N)_{\rm W1}} = \frac{140.3+95.7}{208.5+73.9} \sqrt{1.63\times1.23\times0.74} = 1.02
\end{equation}
i.e. we expect a significance of $2.5\sigma$ in case of a DM signal similar to
the one of LK09.

Fitting a \texttt{powerlaw} to the two spectra simultaneously over the
interval $2.1-10\keV$ gives an index of $1.54^{+0.35}_{-0.18}$ and a
normalisation of $(1.22^{+0.74}_{-0.29})\times 10^{-6}$
photons keV$^{-1}$ cm$^{-2}$ arcmin$^{-2}$. The fit is excellent with $\chi=1135$ for 1078
d.o.f. (reduced $\chi^2$=1.05).

The ratio $(M^{fov}_{DM} / D_L^2)_{Scu} = 1138 M_\odot/\kpc^2$ for Sculptor
which is only 0.605 of the same ratio for Willman~1 (including the Milky Way
contribution in both cases) giving a flux normalization of $2.03\times10^{-6}
\mathrm{photons\, cm^{-2}s^{-1}}$. We add this Gaussian to the power law model
varying the position over the energy interval 2.1-2.8~keV. The quality of the
fit \emph{worsen} (instead of expected improvement) with the lowest increase
in $\chi^2$ being 1141 at $2.3$~keV. At $2.5~\keV$ the increased value is 1144
which allows us to exclude the possibility of a $2.5\keV$ feature to be a
decay line at the level of $\sqrt{1144-1135}=3\sigma$.

\section{Discussion}
\label{sec:discussion}

   \begin{table*}
   \centering
   \begin{tabular}[c]{lcccccc}
     \hline
     Observations & Powerlaw index, & PL normalization & Line
     position & Significance & Line flux & Line flux \\
     & best-fit value & best-fit value, & [keV] & & best-fit value & $3\sigma$ \emph{upper} bound \\
     &  & $\mathrm{10^{-6} ph./(cm^{2}~s~arcmin^{2}}\keV)$ &  & & $\mathrm{ph.
       /(cm^{2}~s~arcmin^{2})}$ & $\mathrm{ph./(cm^{2}~s~arcmin^{2})}$ \\
     \hline
     \texttt{M31on} & 1.32 & 1.66 / 1.82 / 2.39 & 2.55 & $1.7\sigma$ & $7.35\times 10^{-9}$ & $2.09\times
     10^{-8}$\\
     \texttt{M31off} & 1.42 & 1.51 / 1.19 / 2.22 & 2.47 & $2.3\sigma$ & $6.69\times 10^{-9}$ & $3.10\times
     10^{-8}$\\
     \texttt{M31out} & \ldots & \ldots / \ldots / \ldots & 2.52 & $1.6\sigma$ & $3.76\times 10^{-9}$ & $1.16\times
     10^{-8}$\\
     \texttt{Fornax} & 1.37 & 0.79 / 0.59 / 0.70  & 2.19 & $2.6\sigma$ & $9.63\times 10^{-9}$ &
     $2.11\times 10^{-8}$\\
     \hline
   \end{tabular}
   \caption{Parameters of the best-fit \texttt{powerlaw} components (assuming the
     \texttt{powerlaw+gauss} model) and of the maximally allowed flux in the narrow
     Gaussian line in the interval $2.1-2.8$~keV. The slight
     difference between these values and those of
     Table~\ref{tab:best_fit_values} is due to the presence of
     \texttt{gaussian} component and the proportionality of \texttt{powerlaw}
     normalization to field-of-view solid angle (for \texttt{M31on}); for each
     observation, the corresponding parameters coincide within 90\% confidence
     range. The values of best-fit \texttt{powerlaw} parameters for
     \texttt{M31out} are not shown because they are different for different
     observations (see Table~\ref{tab:best_fit_values}).  }
   \label{tab:3sigma}
  \end{table*}

  This work provides more than $80$ times improvement of statistics of
  observations, as compared to LK09 (factor $\sim 5$ times larger total
  exposure time and factor $\sim 4$ improvement in both the effective area and
  the field-of-view). Even under the most conservative assumption about the DM
  column density such an improvement should have led to an about $ 14\sigma$
  detection of the DM decay line. In our analysis no significant lines in the
  position, predicted by LK09 were found.  However, in the \xmm\ observations
  that we processed there are several spectral features in the range $2-3\kev$
  (see Table~\ref{tab:3sigma}). This is not surprising, as the \xmm\
  gold-coated mirrors have Au absorption edge at $\sim 2.21\kev$
  \citep{CAL-TN-0018-2-0,Kirsch:04} and therefore their effective areas
  possess several prominent features in the energy range $2-3\kev$.  According
  to \xmm calibration report \citep{CAL-TN-0018-2-0,Kirsch:04}, there is about
  5\% uncertainty at the modeling of the effective area of both MOS and PN
  cameras at these energies.  These uncertainties in the effective area can
  lead to artificial spectral features due to the interaction of the satellite
  with cosmic rays.\footnote{See e.g.
    \url{http://www.star.le.ac.uk/~amr30/BG/BGTable.html}.}  In particular,
  the solar protons with energies of few hundreds keV can be interpreted as
  X-ray photons (so called \emph{soft proton flares}).  The interaction
  efficiency of solar protons with the instrument is known to be totally
  different from that of real photons~\citep{Kuntz:08}.  In particular, their
  flux is not affected by the Au edge of the \xmm mirrors.  According
  to~\cite{Kuntz:08}, the spectrum of soft proton flares for EPIC cameras is
  well described by the \emph{unfolded} (i.e. assuming that the instrument's
  response is energy-independent) broken powerlaw model with the break energy
  around $\sim 3.2$~keV, \texttt{bknpow/b} in \texttt{XSPEC v.11}. Therefore,
  modeling the spectrum that is significantly contaminated with soft proton
  flares in a standard way (i.e. by using \emph{folded} \texttt{powerlaw}
  model) will produce artificial residual excess at energies of sharp
  decreases of the instrument's effective area.

  In our data analysis we performed both the standard flare
  screening~\citep{Read:03,Nevalainen:05,Carter:07} that uses the inspection
  of high-energy (10--15~keV) light curve and an additional soft proton flare
  cleaning.  As found by~\cite{Pradas:05}, it is possible to screen out the
  remaining flares, e.g. by the visual inspection of the cleaned lightcurve at
  low and intermediate energies. To provide the additional cleaning after the
  rejection of proton flares at high energies, we followed the procedure
  of~\citet{DeLuca:03,Kuntz:08}, leaving only the time intervals, where the
  total count rate differs from its mean value by less than $2\sigma$.

\begin{figure*}
\centering
\includegraphics[width=0.35\textwidth,angle=270,clip=true]{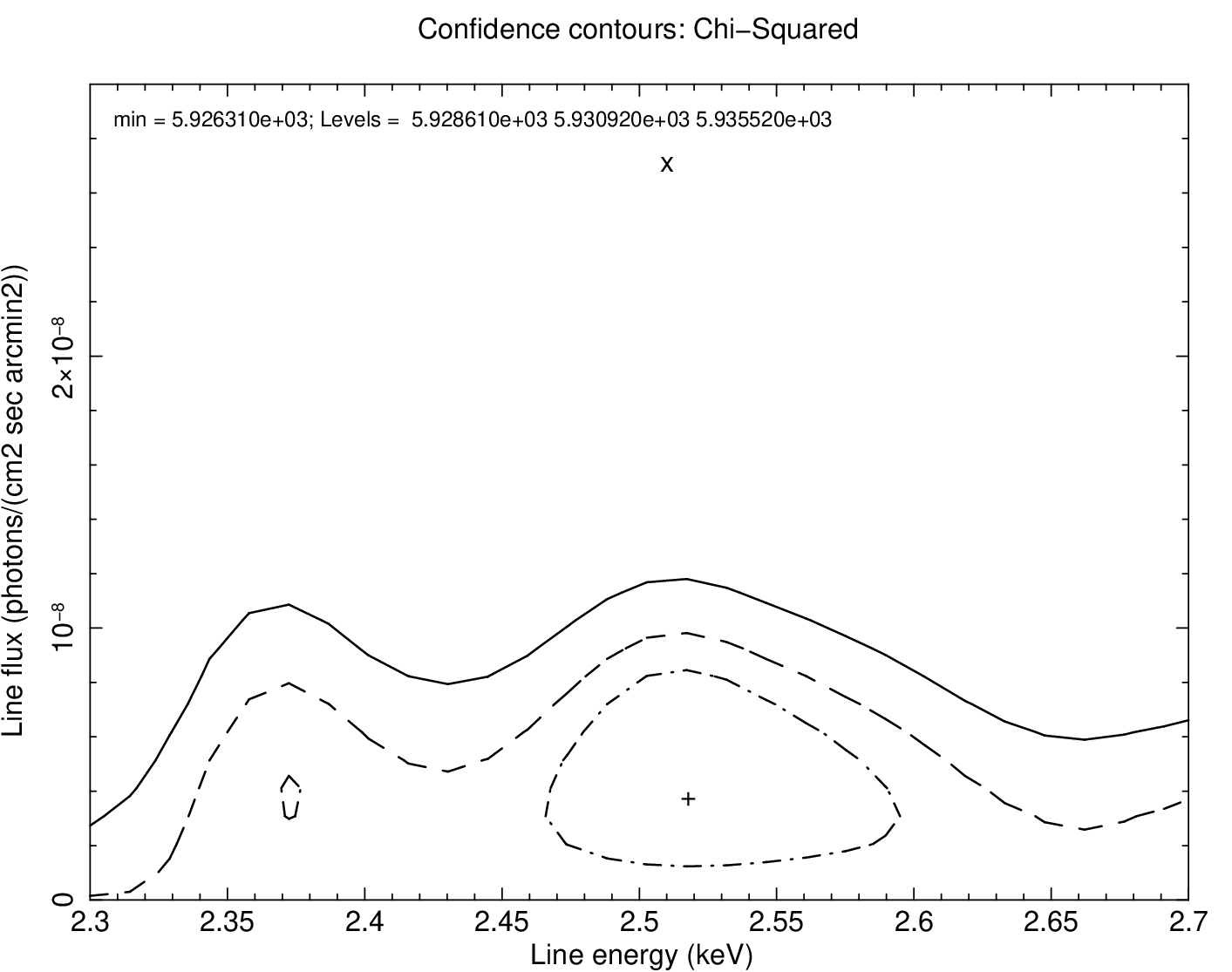}
\includegraphics[width=0.35\textwidth,angle=270,clip=true]{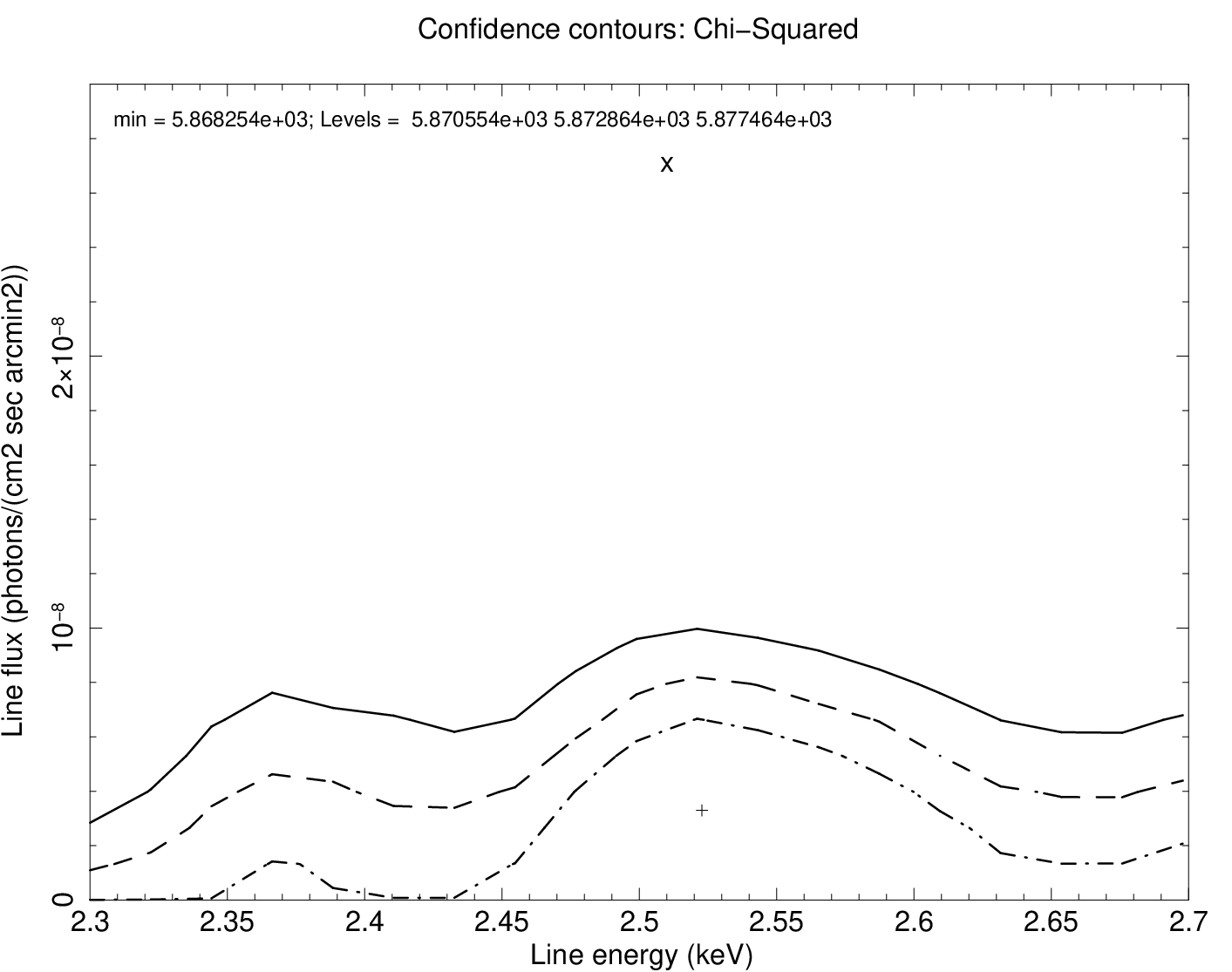}
\caption{\textit{Left:} $\Delta \chi^2 = 2.3$ (dashed-dotted), $4.61$ (dashed)
  and $9.21$ (solid) contours (corresponding to 68\%, 90\% and 99\% confidence
  intervals) for a thin \texttt{gaussian} line allowed by the joint fit of
  \texttt{M31out} spectra used in this work. The resulting line normalizations
  are shown for an averaged DM column density in the \texttt{M31out} region
  (about $170 M_\odot\pc^{-2}$, including the Milky Way contribution). The
  rescaled LK09 line is marked with a cross \texttt{x}; its normalization is
  therefore $4.50\times 10^{-8}\times \frac{170}{208.5+73.9} = 2.71 \times
  10^{-8} \mathrm{photons}\, \cm^{-2} \s\; \mathrm{arcmin}^{-2}$.  It is
  clearly seen that the hypothesis of DM origin for a LK09 line-like feature
  is excluded at extremely high significance, corresponding to more than
  $\approx 10\sigma$ (see text). \textit{Right:} The same as in the left
  figure but with an addition of a \texttt{bknpow/b} component to model the
  contribution of the remaining soft proton contamination (see text). The
  break energy is fixed as 3.2~keV so no degeneracy with the parameters of the
  \texttt{gaussian} is expected. The \texttt{bknpow/b} powerlaw indices at low
  and high energies are fixed to be the same for different cameras observing
  the same spatial region. The line significance drops below $1\sigma$ in this
  case.}
\label{fig:m31out}
\end{figure*}

To test the possible instrumental origin of the features, discussed in this
Section, we first performed only a high-energy light curve
cleaning~\citep{Read:03,Nevalainen:05,Carter:07} and determined the maximally
allowed flux in a narrow line in the energy interval of interest.  After that
we performed an additional soft proton cleaning. This procedure improved the
quality of fit. The significance contours of the resulting ``line'' (for
\texttt{M31out} region) are shown on the left panel in Fig.~\ref{fig:m31out}. In deriving these limits we allowed both the parameters of the background, the
position of the line and its total normalization to vary (the latter from
negative values). Finally, we added an unfolded broken powerlaw component \texttt{bknpow/b} to
model a contribution of remaining soft proton flares (see
e.g.~\citealt{Kuntz:08}). The break energy was fixed as 3.2~keV so no
degeneracy with the parameters of the \texttt{gaussian} is expected. The
\texttt{bknpow/b} powerlaw indices at low and high energies were fixed to be
the same for different cameras observing the same spatial region.  further
improved the quality of fit and made the line detection totally insignificant
(see right Fig.~\ref{fig:m31out} for details).  Our analysis clearly suggests
the instrumental origin of the line-like features at the energy range
2.1--2.8~keV.

As a final note, we should emphasize that the \Chandra\ ACIS instrument used
in LK09 has the Ir absorption edges near 2.11 and 2.55~keV
(see~\citealt{Mirabal:10} for additional discussion). They also report
negative result of searching for the LK09 feature, in agreement with our
findings.

\section{Conclusion}
\label{sec:conclusion}

In this work we demonstrated that the DM decay line can be unambiguously
distinguished from spectral features of other origin (astrophysical or
instrumental) by studying its spacial distribution.  Many DM-containing
objects would provide a comparable DM decay
signal~\citep{Boyarsky:06c,Bertone:07,Boyarsky:09b} which makes such a study
possible~\citep[cf.][]{Boyarsky:09a,denHerder:09,Riemer:09a}. If an
interesting candidate line is found in an object with an unusually high column
density, the differences (always within an order of magnitude) in column
densities with other objects can be compensated by longer exposure or bigger
grasp of observing instruments.

To illustrate the power of this strategy, we have applied this approach to the
recent result of LK09 who put forward a hypothesis about a decaying DM origin
of a $\sim 2.51\keV$ spectral feature in Chandra observations of the Milky Way
satellite known as Willman~1.  Although the parameters of decaying DM,
corresponding to such an interpretation, would lie in the region, already
excluded at least $3\sigma$ level by several existing
works~\citep{Watson:06,Abazajian:06b,Boyarsky:07a} we performed a new
dedicated analysis of several archival observations of M31, Fornax and
Sculptor dSphs.

The exclusions provided by M31 are extremely strong (at the level of
$10-26\sigma$, depending on the observed region). The $10\sigma$ bound is
obtained using the off-center ($10-20$ kpc) observations and a very
conservative model of DM distribution with a heavy stellar disc and a very
large ($\sim 28\kpc$) DM core (the model of~\cite{Corbelli:09}, see
Section~\ref{sec:dm-m31}). Such a model leaves DM column density almost
unchanged from about $\sim 30\kpc$ inward.  Combining all \xmm\ observations
of M31 at the distances $1$--$20\kpc$ off-center provides more than $13\sigma$
bound assuming the model by~\cite{Corbelli:09} and the $26\sigma$ bound with
the model M31B of~\cite{Widrow:05} (with more physically motivated values of
the mass of the stellar disk of M31).

Fornax dSph provides an exclusion between $3.3\sigma$ an $ 5\sigma$ (depending
on how far from its nominal value $2.51\pm 0.07$~keV the position of the line
is allowed to shift) instead of $\sim 4\sigma$ detection expected if the
spectral feature of LK09 was of the DM origin. The bound of $3.3\sigma$
corresponds to the interval $2.3-2.72$ keV and the $5\sigma$ exclusion
corresponds to $2.44-2.58$ keV interval.

\emph{To summarize,} by comparing the strength and position of the feature
found by Loewenstein \& Kusenko with observations of several DM-dominated
objects (M31, Fornax and Sculptor dSphs) we found that the hypothesis of dark
matter origin of LK09 is excluded with the combined significance exceeding
$14\sigma$ even under the most conservative assumptions. This is possible
because of the large increase of the statistics in our observations as
compared with the observation used in LK09. 

To change this conclusion, a number of extreme (and not otherwise motivated)
systematic errors would need to be present in determination of DM content of
M31, Fornax and Sculptor dSph. The DM origin of the spectral feature of LK09
would remain consistent with existing archival data only if the DM amount in
M31, Fornax and Sculptor are strongly overestimated and/or the mass of
Willman~1 is grossly underestimated even compared to the best-fit model
of~\citet{Strigari:07a}.  It should be stressed that in our calculations of
the expected signal in other objects, we have adopted for Willman~1 the DM
column density that results from the mass modeling of \citet{Strigari:07a}.
Our own modeling indicates that the mass of Willman~1 is highly uncertain.
Even supposing that the measured velocity dispersion of Willman~1 provides a
clean estimate of the mass, the general models of \citet{Walker:09} allow the
mass of Willman~1 to vary over several orders of magnitude within the region
of interest.  Furthermore, plausible scenarios such as tidal heating and/or
contamination from unresolved binaries in the kinematic sample would cause the
Willman~1 mass to be over-estimated.  We conclude that the column density
derived from the modeling of \cite{Strigari:07a} is approximately an upper
limit.  In contrast, the large kinematic samples of Fornax and Sculptor
provide tighter mass constraints under the assumptions of our modeling, and
given the larger inherent velocity dispersion, the modeling assumptions are
more secure for Fornax and Sculptor than they are for Willman~1.  As a result,
we expect that significance of our exclusion of dark matter origin of LK09
feature to be very conservative.

Small line-like features (with the significance $\sim 2\sigma$) are present in
the \xmm spectrum at energies between 2 and 3~keV (see Fig.~\ref{fig:m31out},
left and Table~\ref{tab:3sigma} for details).  Their most probable cause is
the interaction of the satellite with soft protons (see discussion in
Section~\ref{sec:discussion}) which can lead to an appearance of artificial
spectral features due to the uncertainties in the effective area at these
energies (c.f.~\citealt{CAL-TN-0018-2-0,Kirsch:04}). Indeed, the significance
of the detection of these features drops below $1\sigma$ when the residual
contamination with soft protons is modeled by an unfolded broken powerlaw (see
previous Section for details).


The search for decaying DM is a well-motivated task, having crucial importance
not only for cosmology and astrophysics, but also for particle physics.  X-ray
astronomy has a significant potential in this
respect~\citep{Abazajian:09a,denHerder:09}.  Due to the small amount of
baryons present in dwarf spheroidal galaxies, the uncertainties in DM
determinations are smaller (as compared to e.g. spiral galaxies) and DM
modeling provides tight bounds on the amount of DM.  However, the
above-mentioned uncertainties make Willman~1 (or other ultra-faint dwarfs) to
be bad observational targets from the point of view of searching for decaying
DM. On the contrary, the ``classical'' dSphs provide excellent observational
targets~\citep{Boyarsky:06c}.  A signal in Fornax, Sculptor (or other
``classical'' dSphs such as Draco or Ursa Minor not considered here because of
the absence of archival observations for them) would therefore have the
potential to provide much more information about the nature of DM than would a
signal from Willman~1.  To improve significantly the best available bounds
\cite[see e.g.][]{Boyarsky:09a} and to probe the most interesting region of
parameter space, an extended (300-500 ks) observations of the well-studied
dSphs are needed.  However as we are searching for the weak diffuse signal, a
special care should be taken in choosing the instruments and observational
times in such a way that minimize possible flare contamination \citep[see
e.g.][]{Boyarsky:06d}.  Drastic improvement in the decaying DM search is
possible with a new generation of spectrometers, having large field of view
and energy resolution close to several
eV~\citep{Boyarsky:06f,Piro:08,denHerder:09,Abazajian:09a}.

\subsubsection*{Acknowledgments.}  We are grateful to L.~Chemin, E.~Corbelli,
M.~Markevitch, S.~Molendi, J.~Nevalainen, M.~Shaposhnikov for many useful
comments and discussions.  DI is also grateful to Scientific and Educational
Centre of the Bogolyubov Institute for Theoretical Physics in Kiev, Ukraine,
and especially to V.~Shadura, for creating wonderful atmosphere for young
Ukrainian scientists. The work of AB and OR was supported in part by the Swiss
National Science Foundation. The work of DI is supported in part from the
SCOPES project No.~IZ73Z0\_128040 of Swiss National Science Foundation, the
Cosmomicrophysics programme and the Program of Fundamental Research of the
Physics and Astronomy Division of the National Academy of Sciences of Ukraine.
The Dark Cosmology Centre is funded by the Danish National Research
Foundation.


\let\jnlstyle=\rm\def\jref#1{{\jnlstyle#1}}\def\aj{\jref{AJ}}
  \def\araa{\jref{ARA\&A}} \def\apj{\jref{ApJ}\ } \def\apjl{\jref{ApJ}\ }
  \def\apjs{\jref{ApJS}} \def\ao{\jref{Appl.~Opt.}} \def\apss{\jref{Ap\&SS}}
  \def\aap{\jref{A\&A}} \def\aapr{\jref{A\&A~Rev.}} \def\aaps{\jref{A\&AS}}
  \def\azh{\jref{AZh}} \def\baas{\jref{BAAS}} \def\jrasc{\jref{JRASC}}
  \def\memras{\jref{MmRAS}} \def\mnras{\jref{MNRAS}\ }
  \def\pra{\jref{Phys.~Rev.~A}\ } \def\prb{\jref{Phys.~Rev.~B}\ }
  \def\prc{\jref{Phys.~Rev.~C}\ } \def\prd{\jref{Phys.~Rev.~D}\ }
  \def\pre{\jref{Phys.~Rev.~E}} \def\prl{\jref{Phys.~Rev.~Lett.}}
  \def\pasp{\jref{PASP}} \def\pasj{\jref{PASJ}} \def\qjras{\jref{QJRAS}}
  \def\skytel{\jref{S\&T}} \def\solphys{\jref{Sol.~Phys.}}
  \def\sovast{\jref{Soviet~Ast.}} \def\ssr{\jref{Space~Sci.~Rev.}}
  \def\zap{\jref{ZAp}} \def\nat{\jref{Nature}\ } \def\iaucirc{\jref{IAU~Circ.}}
  \def\aplett{\jref{Astrophys.~Lett.}}
  \def\apspr{\jref{Astrophys.~Space~Phys.~Res.}}
  \def\bain{\jref{Bull.~Astron.~Inst.~Netherlands}}
  \def\fcp{\jref{Fund.~Cosmic~Phys.}} \def\gca{\jref{Geochim.~Cosmochim.~Acta}}
  \def\grl{\jref{Geophys.~Res.~Lett.}} \def\jcp{\jref{J.~Chem.~Phys.}}
  \def\jgr{\jref{J.~Geophys.~Res.}}
  \def\jqsrt{\jref{J.~Quant.~Spec.~Radiat.~Transf.}}
  \def\memsai{\jref{Mem.~Soc.~Astron.~Italiana}}
  \def\nphysa{\jref{Nucl.~Phys.~A}} \def\physrep{\jref{Phys.~Rep.}}
  \def\physscr{\jref{Phys.~Scr}} \def\planss{\jref{Planet.~Space~Sci.}}
  \def\procspie{\jref{Proc.~SPIE}} \let\astap=\aap \let\apjlett=\apjl
  \let\apjsupp=\apjs \let\applopt=\ao

\end{document}